\documentclass[fleqn,10pt]{wlscirep}
\usepackage[utf8]{inputenc}
\usepackage[T1]{fontenc}
\usepackage[acronym]{glossaries-extra}
\newcommand{\doiprefix}{}

\title{\acrshort{wntrac}: AI Assisted Tracking of Non-pharmaceutical Interventions Implemented Worldwide for \acrshort{covid}}

\author[1,*]{Parthasarathy Suryanarayanan}
\author[1]{Ching-Huei Tsou}
\author[1]{Ananya Poddar}
\author[1]{Diwakar Mahajan}
\author[1]{Bharath Dandala}
\author[2]{Piyush Madan}
\author[1]{Anshul Agrawal}
\author[3]{Charles Wachira}
\author[3]{Osebe Mogaka Samuel}
\author[4]{Osnat Bar-Shira}
\author[3]{Clifton Kipchirchir}
\author[3]{Sharon Okwako}
\author[3]{William Ogallo}
\author[3]{Fred Otieno}
\author[3]{Timothy Nyota}
\author[3]{Fiona Matu}
\author[4]{Vesna Resende Barros}
\author[4]{Daniel Shats}
\author[4]{Oren Kagan}
\author[3]{Sekou Remy}
\author[3]{Oliver Bent}
\author[3]{Pooja Guhan}
\author[1]{Shilpa Mahatma}
\author[3]{Aisha Walcott-Bryant}
\author[1]{Divya Pathak}
\author[4]{Michal Rosen-Zvi}
\affil[1]{IBM Research, Yorktown Heights, USA}
\affil[2]{IBM Research, Cambridge, USA}
\affil[3]{IBM Research, Nairobi, Kenya}
\affil[4]{IBM Research, Mount Carmel Haifa, Israel}
\affil[*]{corresponding author(s): Parthasarathy Suryanarayanan (psuryan@us.ibm.com)}

\usepackage{enumitem}
\usepackage[scaled=.8]{beramono} 

\usepackage{tabularx}
\usepackage{multirow}
\usepackage{cellspace}

\usepackage{subcaption}
\usepackage{wrapfig}
\usepackage{url}
\glsdisablehyper

\usepackage{amsmath}

\usepackage{etoolbox}
\providetoggle{inplace}
\settoggle{inplace}{true}

\glssetcategoryattribute{general}{glossnamefont}{emph}  
\glssetcategoryattribute{general}{textformat}{emph}
\glssetcategoryattribute{acronym}{glossnamefont}{emph}
\glssetcategoryattribute{component}{textformat}{textbf}
\glssetcategoryattribute{data_structure}{textformat}{emph}
\newabbreviation{nlp}{NLP}{natural language processing}
\newabbreviation{ai}{AI}{artificial intelligence}
\newabbreviation{pos}{POS}{part-of-speech}
\newabbreviation{json}{JSON}{JavaScript Object Notation}
\newabbreviation{ml}{ML}{machine learning}
\newabbreviation{iaa}{IAA}{inter-annotator agreement}

\newglossaryentry{event}{name=event, description={An \gls{npi} event}, category={data_structure}}
\newglossaryentry{evidence}{name=evidence, description={A text span supporting the \gls{npi} event}, category={data_structure}}
\newglossaryentry{document}{name=document,description={Document}, category={data_structure}}
\newglossaryentry{type}{name=type,description={type}, category={data_structure}}
\newglossaryentry{value}{name=value,description={value}, category={data_structure}}
\newglossaryentry{restriction}{name=restriction,description={restriction}, category={data_structure}}
\newglossaryentry{npi-index}{name=\gls{npi} Index, description={npi index}, category={data_structure}}

\newglossaryentry{db}{name=database,description={Database}, category={component}}
\newglossaryentry{ui}{name=UI,description={User Interface}, category={component}}
\newglossaryentry{tool}{name=\gls{wntrac} Curator,description={Validation Tool}, category={component}}
\newglossaryentry{browser}{name=\gls{wntrac} data browser,description={Data browser tool}, category={component}}
\newglossaryentry{crawler}{name=crawler,description={Crawler}, category={component}}
\newglossaryentry{prep}{name=Pre-processing,description={Pre-Processing}, category={component}}
\newglossaryentry{sent-class}{name=Sentence classification,description={Sentence classification}, category={component}}
\newglossaryentry{ner-d}{name=Named entity recognition and named entity disambiguation, description={Named entity recognition and named entity disambiguation}, category={component}}
\newglossaryentry{val-ex}{name=Value extraction,description={Value extraction}, category={component}}
\newglossaryentry{delta}{name=change detection,description={change detection}, category={component}}
\newglossaryentry{npi-ext}{name=\gls{npi} extraction, description={\gls{npi} extraction}, category={component}}

\newabbreviation{covid}{COVID-19}{Coronavirus disease 2019}
\newabbreviation{wntrac}{WNTRAC}{Worldwide Non-pharmaceutical Interventions Tracker for COVID-19}
\newabbreviation{npi}{NPI}{non-pharmaceutical intervention}
\newabbreviation{lstm}{LSTM}{long short-term memory}
\newabbreviation{rnn}{RNN}{recurrent neural network}
\newabbreviation{cnn}{CNN}{convolutional neural network}
\newabbreviation{gru}{GRU}{gated recurrent unit}
\newabbreviation{bert}{BERT}{Bidirectional Encoder Representations from Transformers}
\newabbreviation{oxcgrt}{OxCGRT}{Oxford COVID-19 Government Response Tracker}

\glsresetall
\begin{abstract}
The \gls{covid} global pandemic has transformed almost every facet of human society throughout the world. Against an emerging, highly transmissible disease with no definitive treatment or vaccine, governments worldwide have implemented \gls{npi} to slow the spread of the virus. Examples of such interventions include community actions (e.g. school closures, restrictions on mass gatherings), individual actions (e.g. mask wearing, self-quarantine), and environmental actions (e.g. public facility cleaning). We present the \gls{wntrac}, a comprehensive dataset consisting of over 6,000 \glspl{npi} implemented worldwide since the start of the pandemic. \gls{wntrac} covers \glspl{npi} implemented across 261 countries and territories, and classifies \gls{npi} measures into a taxonomy of sixteen \gls{npi} types. \gls{npi} measures are automatically extracted daily from Wikipedia articles using natural language processing techniques and manually validated to ensure accuracy and veracity. We hope that the dataset is valuable for policymakers, public health leaders, and researchers in modeling and analysis efforts for controlling the spread of \gls{covid}.

 \end{abstract}

\begin{document}
\flushbottom
\maketitle
\thispagestyle{empty}

\glsresetall
\section*{Background \& Summary}\label{sec:background}
\iffalse
    1. Say something about the COVID19  - DONE
    2. relevance of the API in controlling or easing COVID-19 - DONE
    3. varieties of NPI - DONE
    4. talk about how an NPI dataset might help - DONE
    6. talk about the existing datasets and their problems / desirable characteristics of an NPI dataset - DONE
    7. Justify our approach (Wikipedia-based) - DONE
    8. introduce our dataset formally - DONE
    9. Set expectation on what is lies ahead in the paper - DONE
\fi

The \gls{covid} pandemic has made an unprecedented impact on almost every facet of human civilization from healthcare systems, to economies and governments worldwide. As of August 2020, every country in the world has been affected, with more than 24M confirmed cases of infection and death toll approaching a million cases worldwide~\cite{jhu, who, worldometer}. The pandemic has triggered a wide range of \gls{npi} responses across the world. With therapeutic and preventive interventions still in early stages of development, every country has resorted to \gls{npi} as a primary strategy~\cite{c19hcc, ferguson2020report} for disease control. Examples of such interventions include community actions (e.g. school closures, restrictions on mass gatherings), individual actions (e.g. mask wearing, self-quarantine), and environmental actions (e.g. public facility cleaning). Such \glspl{npi} vary significantly in their implementation based on the maturity of the health infrastructure, robustness of the economy and cultural values unique to the region. 

Public health policy makers worldwide are striving to introduce successful intervention plans to manage the spread of disease while balancing the socio-economic impacts~\cite{coibion2020cost, lancet2020india}. These initiatives will benefit from modeling the efficacy of different intervention strategies. The pandemic has sparked an ongoing surge of discovery and information sharing resulting in an unprecedented amount of data being published online~\cite{Wang2020CORD19TC}. This includes information about \gls{npi} measures, which are available in a wide variety of unstructured data sources, including official government websites~\cite{us-chamber-of-commerce,us-csg}, press releases, social media, and news articles. However such modeling requires the information about the \glspl{npi} to be available in a structured form. 

To address this urgent need, several data collection initiatives have emerged in the recent months resulting in several publicly available datasets with varying degrees of coverage, data freshness, and sparsity. For example, the CoronaNet dataset~\cite{CoronaNet} contains the monadic and dyadic data on policy actions taken by governments across the world, manually curated by over 500 researchers covering sixteen \gls{npi} types and is kept fairly up-to-date. The Complexity Science Hub, Vienna enlisted researchers, students and volunteers to curate the \emph{Complexity Science Hub COVID-19 Control Strategies List}~\cite{Desvars-Larrive2020} dataset, of eight different \gls{npi} types but covering only 57 countries. Similarly, the Oxford \gls{covid} Government Response Tracker~\cite{hale2020oxford} dataset, takes a crowd-sourcing approach and covers 17 \gls{npi} types, 186 regions, 52 US states and territories. Because all these datasets are assembled manually, each of them is constrained in one or more respects: geographical scope, taxonomic richness, frequency of updates or granularity of details, and evidential sources. An AI-assisted, semi-automated data collection approach, driven by a rich, extensible taxonomy, can help overcome these issues and may result in a larger, frequently updated dataset with less manual labor. 

\iftoggle{inplace}{
    \begin{figure}[htp!]
  \centering
  \includegraphics[width=0.8\linewidth]{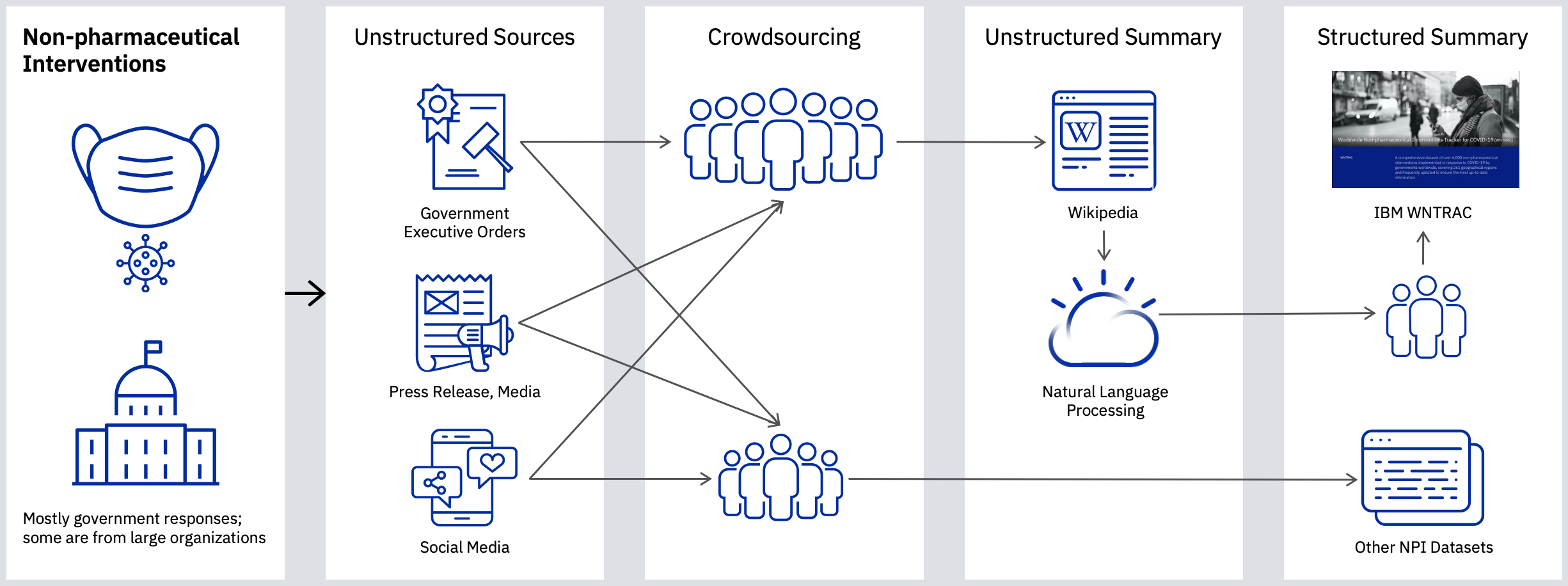}
  \caption{Artificial intelligence assisted approach to build the \gls{wntrac} dataset.}  
  \label{fig:approach}
\end{figure} }

Wikipedia is one of the main sources of accessible information on the Internet. Since the start of \gls{covid}, a dedicated global network of volunteers has been creating, updating, and translating Wikipedia articles with vital information about the pandemic~\cite{wiki-covid}. Over 5,000 new Wikipedia pages on \gls{covid} have been written by more than 71,000 volunteers since the onset of the pandemic accumulating more than 440M page views by June~2020. Wikipedia articles, even though crowd-sourced, through the process of collective validation~\cite{jessen2012aggregated} and by citations of credible sources such as government websites, scientific literature, and news articles can serve as a reliable source of \gls{npi} data. Further, these Wikipedia articles are constantly updated; have been edited more than 793,000 times as of August~2020 making it both a rich and up-to-date source. Based on this, we postulated that an approach based on automated information extraction from Wikipedia, followed by human validation to ensure accuracy and veracity, would result in a frequently updated dataset with a wider coverage compared to any of the existing datasets. We present the result of our work, \gls{wntrac}, a comprehensive dataset consisting of over 6,000 \glspl{npi} implemented worldwide since the start of the pandemic. \gls{wntrac} covers \glspl{npi} implemented across 261 countries and territories, and classifies \gls{npi} measures into a taxonomy of sixteen \gls{npi} categories. \gls{npi} measures are automatically extracted daily from Wikipedia articles using \gls{nlp} techniques and manually validated to ensure accuracy and veracity.

In what follows, we explain the methods used to create the dataset, outline the challenges and key design choices, describe the format, provide an assessment of its quality and lay out our vision of how this dataset can be used by policy makers, public health leaders, and data scientists and researchers to support modeling and analysis efforts. 

\iffalse
    Add the following aspects in the comparison to other datasets:
    region
    up to date - semi-automatic
    coverage -taxonomy
    fine grained entity
    human in the loop
\fi

%
 
\section*{Methods}\label{sec:methods}
\iffalse
    1. Describe how we collected the raw data. Highlight any issues with the approach. Are we missing some data due to this approach? 
    2. Describe how we process the data. Are we introducing any errors (including that of omission) due to this processing?
    3. Describe the details of the system used for processing.
    3a. Define the NPI event.
    3b. Define the language / technical problem clearly including any sub-problems. Describe common approaches to solving such problems
    3c. Describe various stages of the processing pipeline.
    3d. What are the ML models that are used and how they have been trained. Include any qualitative assessment of the accuracy of the models (because it would affect quality of the dataset).
    4. Describe the validation application and any design choices made there (again, because it would affect quality of the dataset)
    5. Describe how this system is scalable and how it allows us to keep up with the changes to the Wikipedia (this is this is where we support the claims made in the background section).
    6. Link it to the next section (quality assessment / technical validation)
\fi

We built a semi-automated system to construct the dataset and keep it current. The \gls{npi} measures are modeled as \glspl{event} and \glspl{evidence} for information extraction purposes. This is illustrated by a motivating example shown in the Figure~\ref{fig:npi-example-may-15}. Each \gls{event} corresponds to an imposition or lifting of a particular \gls{npi}. An \gls{event} is defined to be a 5-tuple (what, value, where, when, restriction), where
\begin{enumerate} [noitemsep, topsep=0pt]
    \item What: the \emph{type} of \gls{npi} that was imposed or lifted. \glspl{npi} are grouped into sixteen major types. In the example, the type is \emph{school closure}.
    \item Value: sub-category or attribute that further qualifies the \gls{npi} type more specifically. In the example, the associated value is \emph{all schools closed}. A detailed description of each type and the corresponding possible values is shown in Table~\ref{tab:taxonomy}. 
    \item Where: the region (country, territory, province, or state) in which the \gls{npi} measure has been implemented or withdrawn. In this example, there are three distinct regions, namely, \emph{Punjab, Chhattisgarh, Manipur} that are identified and three separate \glspl{event} will be extracted.
    \item When: The date from which the \gls{npi} was imposed or lifted. In the example, the date will be \emph{13 March}, corresponding to the implementation of the \gls{npi}, even if a likely date for the cancellation of the \gls{npi}, \emph{31 March}, is indicated.
    \item Restriction: a flag indicating that the event corresponds to the introduction or withdrawal of the \gls{npi}. It should be noted that the lifting of the \gls{npi} is treated as a separate event. In the example, the restriction type is \emph{imposed}. 
\end{enumerate}

\iftoggle{inplace}{
 \begin{figure}[htp!]
  \centering
  \includegraphics[width=0.6\linewidth]{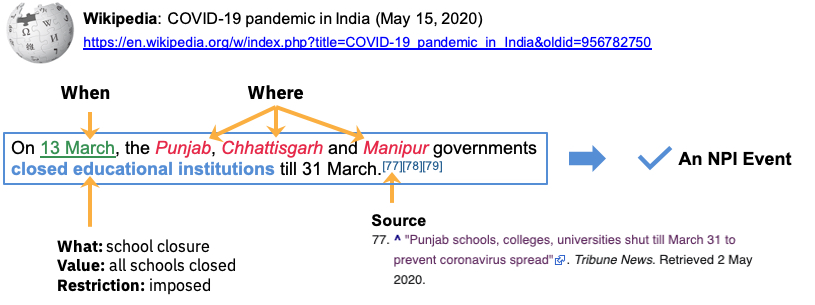}
  \caption{An example of the \gls{npi} measure mentioned in the Wikipedia article of 15\textsuperscript{th} May 2020.}  
  \label{fig:npi-example-may-15}
\end{figure}

\iffalse
    \begin{figure}[htp!]
      \centering
      \includegraphics[width=0.6\linewidth]{wntrac-eg-june-5}
      \caption{An example of the \gls{npi} event reported in the Wikipedia article on 5\textsuperscript{th} June 2020. Compared to the May 15 version of Figure~\ref{fig:npi-example-may-15}, the regions and sources have been updated. }  
      \label{fig:npi-example-june-5}
    \end{figure}
\fi }

In addition to the mandatory fields described above, \gls{event} contains one or more \glspl{evidence}. An \gls{evidence} is a span of text extracted from Wikipedia that discusses a particular \gls{event}. In the example, \emph{On 13 March, the Punjab, Chhattisgarh, and Manipur governments declared holidays in all schools and colleges till 31 March.} is the \gls{evidence}. An \gls{evidence} may support more than one \gls{event}. Each \gls{evidence} is accompanied by a source type indicating the type of source of Wikipedia citation. More details about such additional attributes can be found in the data records section.

\iftoggle{inplace}{
 \begin{figure}[htp!]
  \centering
  \includegraphics[width=0.6\linewidth]{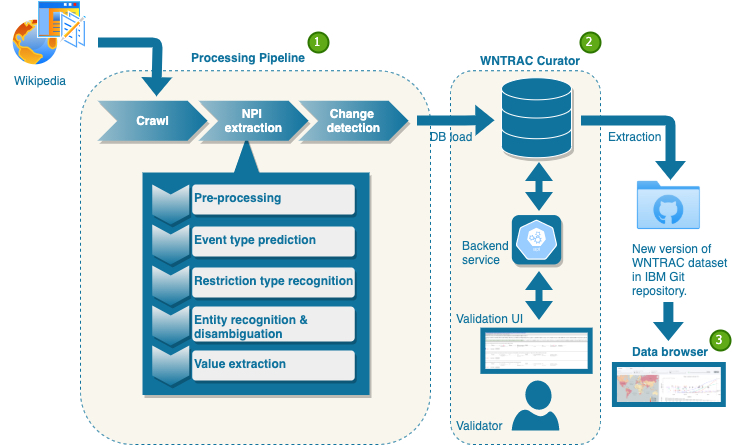}
  \caption{The \gls{wntrac} automated \gls{npi} curation system. It consists of a processing pipeline, \gls{tool} validation tool, and \gls{npi} data browser.}  
  \label{fig:system}
\end{figure} }

The system, shown in the Figure~\ref{fig:system}, is designed to be scalable for continuous gathering, extraction and validation of \gls{npi} \glspl{event}. It consists of three subsystems: a data processing pipeline for capturing and extracting potential \gls{npi} \glspl{event} from Wikipedia articles, a tool called \gls{tool} for human validation of \gls{npi} \glspl{event} automatically extracted using the aforementioned pipeline and a data browser for visualizing the data. In the next section, we describe the system and its components at a high level, focusing on key design choices that have a bearing on the quality of the dataset, starting with a brief description of the data collection.
\iftoggle{inplace}{
    \newcommand{\pf}[1]{\parbox{6cm}{#1}}
\newcommand{\midsepdefault}{\aboverulesep = 0.605mm \belowrulesep = 0.984mm}
\newcommand{\midsepremove}{\aboverulesep = 0mm \belowrulesep = 0mm}

\begin{scriptsize}\centering
\begin{longtable}[t]{@{}p{0.15\textwidth}@{}p{0.35\textwidth}p{0.06\textwidth}@{}m{0.30\textwidth}}
        \toprule
        \textbf{\gls{npi}}
        & \textbf{Example} 
        & \textbf{Value} 
        & \textbf{Value description}\\
        \midrule
 
        {\parbox{2.5cm}{changes in \newline prison-related policies}} 
        & \pf{On March 30, the GNA announced the release of 466 detainees in Tripoli, as part of an effort to stop the spread of the virus in prisons.}
        & Integer
        & Number of prisoners that were released \\
        \midrule
        
        {confinement}
        &\pf{On 19 March, President Alberto Fernández announced a mandatory lockdown to curb the spread of coronavirus.}
        & Category
        & \begin{enumerate}[nosep, noitemsep, leftmargin=*]
            \item  Mandatory/advised for all the population
            \item  Mandatory/advised for people at risk
\end{enumerate} \\
        \midrule
        
        contact tracing
        & \pf{On 2 March, a case in Nimes was traced to the mid-February Mulhouse Megachurch event.}
        & Category 
        & \begin{enumerate}[nosep, noitemsep, leftmargin=*]
            \item Tracing back 14 days of contacts of a confirmed patient through electronic information
            \item Tracing contacts of a person who needs to be isolated as was in contact with a confirmed patient through electronic information
        \end{enumerate} \\
        \midrule
        
        {\parbox{2.5cm} {domestic flight restriction}}	
        & \pf{On 1 April, the Government of Afghanistan suspended flights between Kabul and Herat.}
        & String 
        & Name of the state where the passenger is arriving from \\
        \midrule
        
        economic impact 
        & \pf{Up until 14 March, the Afghan government had spent \$25 million to tackle the outbreak, which included \$7 million of aid packages.}
        & Category  
        & \begin{enumerate}[nosep, noitemsep,  leftmargin=*]
            \item  Stock market
            \item  Unemployment rate
            \item  Industrial production
\end{enumerate} \\
        \midrule
        
        {\parbox{2.5cm} {entertainment / \newline cultural sector closure}}
        & \pf{On April 7, Rockland and Sullivan counties closed their parks.}
        & Category  
        & \begin{enumerate}[nosep, noitemsep,  leftmargin=*]
              \item  Bars, restaurants, night clubs
              \item  Museums, theaters, cinema, libraries, festivities 
              \item  Parks and public gardens
              \item  Gyms and pools 
              \item  Churches
\end{enumerate} \\
        \midrule
        
        {\parbox{2.5cm} {freedom of movement \newline(nationality dependent)}}
        & \pf{Iran was added to the list of countries whose nationals were suspended entry to Cambodia, making a total of six.} 
        & String 
        & Name of the country the citizen is from\\
        \midrule

        {\parbox{2.5cm} {international \newline flight restrictions}}	
        & \pf{With effect from midnight on 1 April, Cuba suspended the arrival of all international flights.}
        & String 
        & Name of the country or state where the passenger is arriving from \\
        \midrule
        
        {\parbox{2.5cm} {introduction of \newline travel quarantine policies}}	
        & \pf{Israeli nationals returning from Egypt were required to enter an immediate 14-day quarantine.}
        & String 
        & Name of the country or state where the passenger travelled from\\
        \midrule
        
        mask wearing 
        & \pf{On April 15, Cuomo signed an executive order requiring all New York State residents to wear face masks or coverings in public places.}
        & Category 
        & \begin{enumerate}[nosep, noitemsep,  leftmargin=*, topsep=0pt] \item Mandatory
            \item Mandatory in some public spaces
            \item Recommended
\end{enumerate} \\
        \midrule
        
        mass gatherings	
        & \pf{On 13 March, it was announced at an official press conference that a four-week ban on public gatherings of more than 100 persons would be put into effect as of Monday 16 March.}
        & Integer 
        & Maximum number of people in social gatherings allowed by the government\\
        \midrule
        
        public services closure
        & \pf{On 19 March, Election Commissioner Mahinda Deshapriya revealed that the 2020 Sri Lankan parliamentary election will be postponed indefinitely until further notice due to the coronavirus pandemic.} 
        & Category 
        & \begin{enumerate}[nosep, noitemsep,  leftmargin=*]
            \item Government/parliament system closed
            \item Legal system closed
\end{enumerate} \\  
        \midrule
        
        public transportation
        & \pf{On March 20, Regina Transit and Saskatoon Transit suspended fares for all bus service, but with reduced service.} 
        & Category 
        &  \begin{enumerate}[nosep, noitemsep,  leftmargin=*]
            \item Partial cancellation of routes/stops during the week/weekend
            \item Total cancellation of transport (special case for some states in China)
\end{enumerate} \\   
        \midrule
        
         school closure
         & \pf{On 13 March, the Punjab and Chhattisgarh governments declared holidays in all schools and colleges till 31 March.}
         & Category 
         & \begin{enumerate}[nosep, noitemsep,  leftmargin=*]
             \item All schools (general) closed
             \item Only kindergartens/daycare closed
             \item Only schools (primary/secondary) closed
             \item Universities closed
\end{enumerate} \\
        \midrule
        
        state of emergency \newline (legal impact)
        & {\pf{Governor Charlie Baker declared a state of emergency for the state of Massachusetts on March 10.}}
        & Category
        & \begin{enumerate}[nosep, noitemsep,  leftmargin=*]
            \item National guard joins the law enforcement
            \item Army joins the law enforcement
\end{enumerate} \\
        \midrule
        
        \renewcommand{\arraystretch}{1.5}work restrictions
        & \pf{On 10 April, Koike announced closure requests for six categories of businesses in Tokyo.}
        & Category 
        & \begin{enumerate}[nosep, noitemsep,  leftmargin=*]
            \item Suggestion to work from home for non-essential workers
            \item Mandatory work from home enforcement for non-essential workers 
\end{enumerate} \\
        \bottomrule
  \caption{Taxonomy of the \acrlong{wntrac} dataset.}\label{tab:taxonomy} 
  \end{longtable}
\end{scriptsize}

}

\subsubsection*{Data Collection}\label{sec:data_collection}
As stated earlier, Wikipedia includes a broad range of articles on \gls{covid} covering a variety of topics, including the cause, transmission, diagnosis, prevention, management, economic impact, and national responses. Categories are used in Wikipedia to link articles under a common topic and are found at the bottom of the article page. This dataset was collected by automatically crawling Wikipedia articles discussing \gls{covid} in different regions belonging to the category~\cite{wiki:catagory} \texttt{\gls{covid} pandemic by country} \footnote{For \emph{mask wearing} \gls{npi} type, Wikipedia articles were observed to be incomplete for some regions, so we augmented the dataset with hand-curated list of \gls{npi} measures from web sources.}. There are 156 subcategories and 198 articles directly under \texttt{\gls{covid} pandemic by country}, and when retrieved recursively, there are 384 articles under this top-level category as of July 2020. Considering the limited availability of volunteers, and the volume of \gls{npi} measures that had to be validated initially, we restricted the number of articles to a manageable size, covering 261 regions (i.e. countries and territories) as listed in the tables at the end of the paper.

\subsubsection*{Processing Pipeline}
The first step in the data processing is to retrieve the aforementioned list of Wikipedia articles on a periodic basis. The \gls{crawler} module implements this functionality. It uses the MediaWiki API~\cite{wiki-api} for downloading the articles. As part of this step, we extract the text content of each article, while at the same time preserving all the associated citations. This process produces a \gls{document} for each article. Each sentence in a \gls{document} is a candidate for \gls{npi-ext}. As of August 2020, the aggregate crawled data contains over 55,000 sentences, with an average of 213 sentences per \gls{document}. The second step in the pipeline is the extraction of the \gls{npi} \glspl{event} from a \gls{document}. It is broken into a sequence of steps described below.

\begin{itemize}
    \item \gls{prep}: As the first step in processing a \gls{document}, we use sentence boundary detection algorithms from libraries such as spaCy~\cite{honnibal2015improved}, to identify where sentences begin and end. Although the sentences are used as logical units to extract \gls{npi} events, we preserved the order in which they appear in the source document for reasons detailed below. Also, at this step, we extract and retain the citation URL, if available for each sentence.
    
    \item \gls{sent-class}: Next, we classify the sentence into one of the \gls{npi} types such as \emph{school closure} to identify potential \gls{npi} \glspl{event}. If no \gls{npi} is discussed in the sentence, we classify it as \emph{discarded}. We use multiple learning algorithms, including logistic regression, Support Vector Machines, and \gls{bert}~\cite{Devlin_Chang_Lee_Toutanova_2018}, and employ an ensemble method to obtain better overall predictive performance. A small subset of the data (1490 sentences), was manually annotated to train the models. Independently, we also categorize the sentence as implying either the introduction or the withdrawal of an \gls{npi} (\gls{restriction}).
    
    \item \gls{ner-d}: After we identify the potential \glspl{event} in the previous step, we extract specific constituent entities for each candidate \gls{event} from the sentence. We used state-of-the-art named-entity recognizers (such as spaCy~\cite{honnibal2015improved}) and normalizers to detect and normalize locations (\emph{Where} : [\emph{Punjab}, \emph{Chattisgarh}, \emph{Manipal} ]) and time expressions (\emph{When} : \emph{March 13}). In addition, we also link the location entities of type ‘GPE’ in the Wikipedia article title to the corresponding ISO codes~\cite{wiki:iso-3166-1, wiki:iso-3166-2}. Even though we use the sentence as a logical unit for the extraction of an \gls{npi} event, the sentence itself may not include all the relevant information. For example, date or location may be available in sentences in the vicinity or in the header of the paragraph to which the sentence belongs. To address this key challenge, we developed a heuristic-based relation detection algorithm to associate one of the extracted dates or locations from the current document to each sentence.
    
    \item \gls{val-ex}: The last step in \gls{npi} event extraction, is determining the associated \gls{value}. We use multiple rule-based algorithms that either operate independently or depend on information extracted by the previous steps. For example, given the sentence \texttt{"On 13 March, it was announced at an official press conference that a four-week ban on public gatherings of more than 100.”}, the event type is \emph{mass gathering} and the associated value is \emph{maximum number of people in social-gathering allowed by the government}. The value extraction is performed using parse-based rule engines~\cite{honnibal2015improved}. It is worth noting that the value extraction components should know the actual type \emph{mass gatherings} before extracting the correct value "100". Similarly, given a sentence \texttt{“On 1 April, the Government of USA suspended flights from New York to Texas”}, the event type is \emph{domestic flight restriction} and the associated value is \emph{name of the state where the passenger is arriving from}. To correctly extract the \gls{value}, the value extraction needs to know the correct \gls{type} and normalized locations ("New York") respectively. 
\end{itemize}

Thus, using the above procedure, we extract the unique 5-tuples that are the candidate \gls{npi} \glspl{event}. Once extracted, they are presented to the volunteers for validation to ensure data quality. This process is repeated every day. In order to minimize manual labor, considering the small number of volunteers, we attempt to detect changes since the last time we crawled Wikipedia. We use a combination of syntactic similarity metrics such as Levenshtein Norm, and semantic similarity metrics such as event attribute matching to perform this daily \gls{delta} for each extracted \gls{document}.

\iftoggle{inplace}{
    \begin{figure}[htp!]
    \centering
    \includegraphics[width=0.99\linewidth]{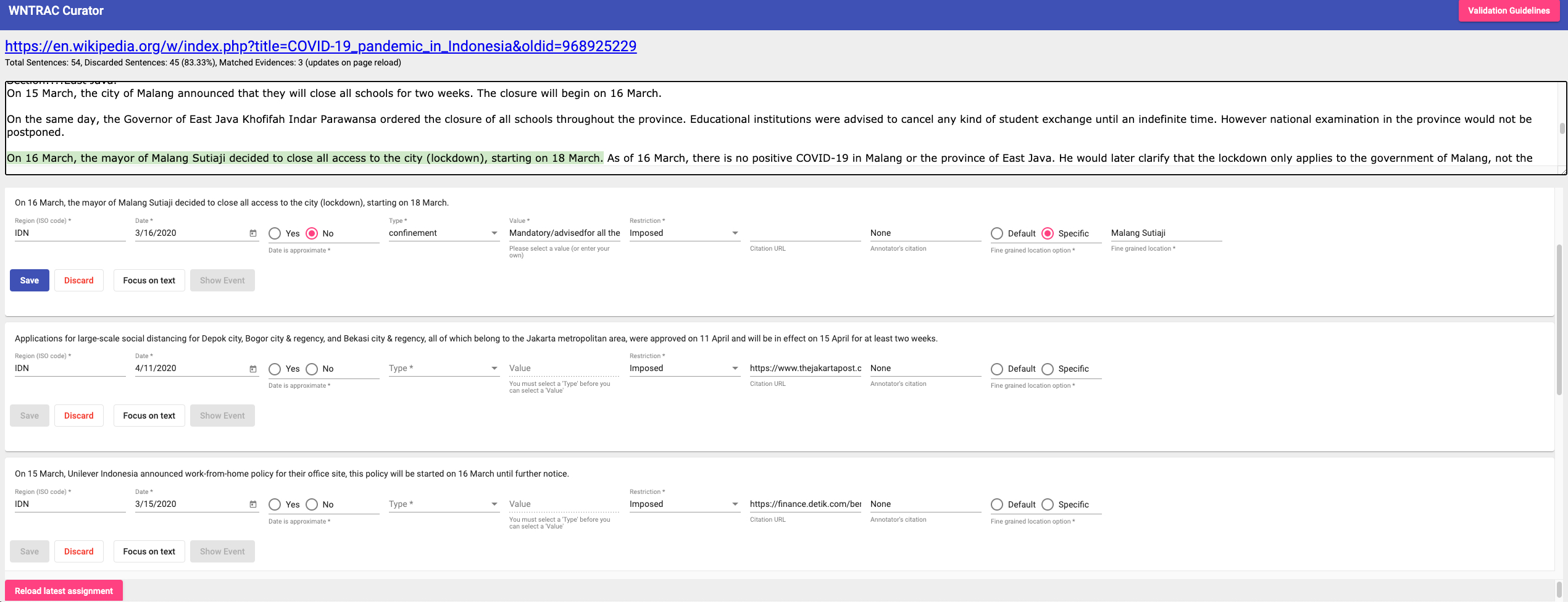}
    \caption{\gls{tool} tool used for ongoing validation of the dataset.}
    \label{fig:wntrac-curator}
\end{figure} }

\subsubsection*{\gls{tool}}
The \glspl{event} automatically extracted from the pipeline are vetted by volunteers using the \gls{tool} validation tool. The tool is a simple web-application backed by a \gls{db} as shown in Figure~\ref{fig:system}. The tool is shown in Figure~\ref{fig:wntrac-curator}. At the top, it displays the complete Wikipedia \gls{document} extracted by the processing pipeline. Below the \gls{document}, each candidate \gls{event} is shown to the volunteer in separate \emph{cards}. The volunteer can adjudge the candidate \gls{event} to be a brand new \gls{npi} event or an \gls{evidence} to an existing \gls{event} or discard the candidate. They can also correct any of the attributes associated with the \gls{event} extracted by the pipeline.        

\iftoggle{inplace}{
    \begin{figure}[htp!]
    \centering
    \includegraphics[width=0.9\linewidth]{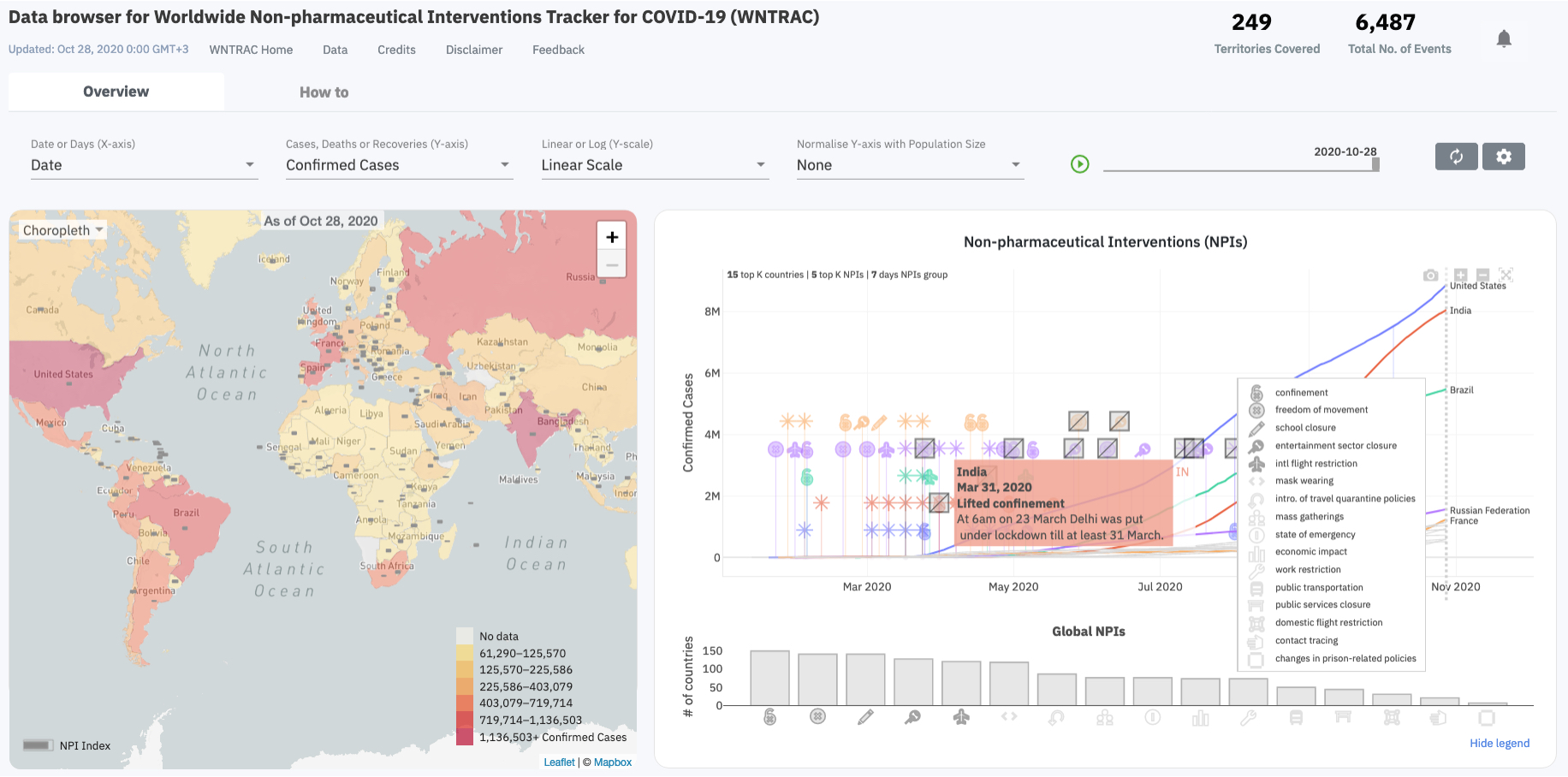}
    \caption{Data browser for visualizing the \acrlong{wntrac} dataset.}
    \label{fig:data_browser}
\end{figure} }

\subsubsection*{Data Browser}
Figure~\ref{fig:data_browser} presents an interactive data browser~\cite{data-browser} that uses a chart, map, and histogram to provide a descriptive analyses of \glspl{npi} and \gls{covid} outcomes such as confirmed cases and deaths. The browser has a control panel used to filter the data being visualized (e.g cases vs deaths), as well as how it is visualized (e.g. linear vs log scale). A play slider can be used to view the temporal evolution of \glspl{npi} and \gls{covid} outcomes in a given region. The chart illustrates the time points in which a geographical region imposes or lifts an \gls{npi} along with the temporal trends of \gls{covid} outcomes. The different types of \glspl{npi} are illustrated using specific icons that are described in a legend. Groups of interventions are noted with the star icon. The number of countries/territories and the number of \glspl{npi} shown in the chart can be adjusted in the settings.  The user can select a specific line on the chart referring to a territory to focus on the \glspl{npi} imposed and lifted in that location. The histogram below the chart shows the number of territories that have imposed the different types of \glspl{npi} and can be selected to see the territories on the map that have imposed the selected subset of \glspl{npi}. The map illustrates the proportion of \gls{npi} categories (out of the 16 \gls{npi} categories in the dataset) implemented in each region using a gray-colored bar. Furthermore, when a region is selected,  the gray-colored bar in any other region illustrates the proportion of \gls{npi} categories in the other region as a proportion of \gls{npi} categories implemented in the selected region. The map is also used to visualize the geographic distribution of the selected \gls{covid} outcome using choropleth, spikes, or bubbles. The user can interact with the territories on the map to focus on a location and view the data on the chart. Note that for some countries such as the United States, the map can be zoomed to reveal finer-grained data for sub-regions such as states.

\section*{Data Records}\label{sec:data-records}

In addition to the key fields discussed earlier, the dataset also contains a few additional attributes for each \gls{event}. A complete listing of all fields across \gls{event} and \gls{evidence} is shown in Table~\ref{tab:data-record}, along with an example for each field. Each version of the dataset consists of two CSV files named \texttt{ibm-wntrac-yyyy-mm-dd-events.csv} and \texttt{ibm-wntrac-yyyy-mm-dd-evidences.csv}, corresponding to \glspl{event} and \glspl{evidence} respectively. A live version of dataset is available in our GitHub repository~\url{https://github.com/IBM/wntrac/tree/master/data} for download. The dataset is regularly updated. At the time of the submission, the dataset is updated as of October 13\textsuperscript{th}, 2020. Historical versions of the dataset are made available in the same GitHub repository. Further, a static copy of the dataset containing \glspl{npi} recorded as of 8\textsuperscript{th} July 2020, used for the technical validation in the paper has been archived in figshare~\cite{figshare}. In the next section, we include some high-level dataset statistics to provide a sense of the distribution of the data.
\iftoggle{inplace}{
    \begin{table*}[htp!]
    \scriptsize
\centering
    \begin{tabular}[t]{p{0.10\textwidth}p{0.50\textwidth}p{0.30\textwidth}}\toprule
            \textbf{Field name}
            & \textbf{Description} 
            & \textbf{Example}
            \\ \midrule
            even\_id & Globally unique identifier~\cite{wiki:uuid} for the particular \gls{npi} & 7db34fd1-d121-479f-9713-af7596a45aa1 \\
            type    &  Type of the \gls{npi} & School closure \\ 
	        country &  Country where the \gls{npi} was implemented. Name in ISO 3166-1 coding~\cite{wiki:iso-3166-1} & USA \\
            state/province &  State or province where the \gls{npi} was implemented. Name in ISO 3166-2 coding~\cite{wiki:iso-3166-2} & Vermont \\
            date & Date when the \gls{npi} comes to effect. It is not the date of announcement & 2020-03-26 \\
            epoch & Unix epoch time~\cite{wiki:epoch} corresponding to the date & 1589749200000.0 \\ 
            value & Value associated with the \gls{npi}. & Refer to Table for details \\
            restriction & Ordinal values representing imposition ($1$) or lifting ($0$) of an \gls{npi} & 0 \\
            sent\_id & Globally unique identifier~\cite{wiki:uuid} for the evidence sentence & d68ea644-24d5-4abf-93b0-dabc1cd3c2eb \\
            doc\_url & Document URL & \url{https://en.wikipedia.org/wiki/COVID-19_pandemic_in_Vermont} \\
            crawl\_id & Globally unique identifier~\cite{wiki:uuid} for the particular crawl in which this evidence sentence was fetched & 2020-05-06\_d0cba9ae-8fda-11ea-b351-069b8ffc8dc8 \\
            crawl\_date & Date of the crawl that fetched this evidence sentence &  2020\-05\-06 \\
            text & Evidence sentence in the document where the \gls{npi} is discussed & On March 26, Governor Scott ordered all schools in Vermont to remain closed for in-person classes for the rest of the academic year \\
            citation\_url & URL cited for the evidence sentence in the source document & \iffalse \url{https://governor.vermont.gov/content/directive-5-continuity-learning-planning-pursuant-eo-01-20} \fi \\
            anno\_provided\_url & Additional citation URL provided by the human volunteer who performed the validation. & \iffalse \url{https://www.vpr.org/post/gov-closes-vermont-schools-rest-academic-year} \fi \\
            fine\_grained\_location & Geographic locations mentioned in the evidence sentence separated by pipeline. & Vermont \\
            source\_type & Wikipedia citation source type indicating government ($G$) or other sources ($O$) & G \\
             \bottomrule
    \end{tabular}
    \caption{Data record for the \acrlong{wntrac} dataset.} \label{tab:data-record}
\end{table*} }

\subsection*{Dataset Statistics}
Figure~\ref{fig:stats-npi-distribution} shows the distribution of the \gls{npi} measures imposed worldwide. \emph{Entertainment / cultural sector closure}, \emph{confinement} and \emph{school closure} are the predominant \glspl{npi} taken by governments\footnote{Figures in Dataset Statistics, Usage Notes sections were generated from the latest version of dataset, dated 13\textsuperscript{th} October 2020, available at the time of manuscript submission. A copy of this version of the dataset is also available in figshare.~\cite{figshare}}. Figure~\ref{fig:stats-region-count-by-npi} summarizes the overall total number of regions that implemented \glspl{npi} of each type. As shown in the graph confinement, \emph{school\ closure} and \emph{freedom of movement} are the most common \glspl{npi} imposed worldwide, as expected from Figure~\ref{fig:stats-npi-distribution}. Figure~\ref{fig:stats-npi-count-by-region} shows the breakdown of the \glspl{npi} within each region, for the top twenty regions that have implemented the highest number of \glspl{npi} measures. 

\iftoggle{inplace}{
    \begin{figure}[htp!]
\centering
    \includegraphics[width=0.5\linewidth]{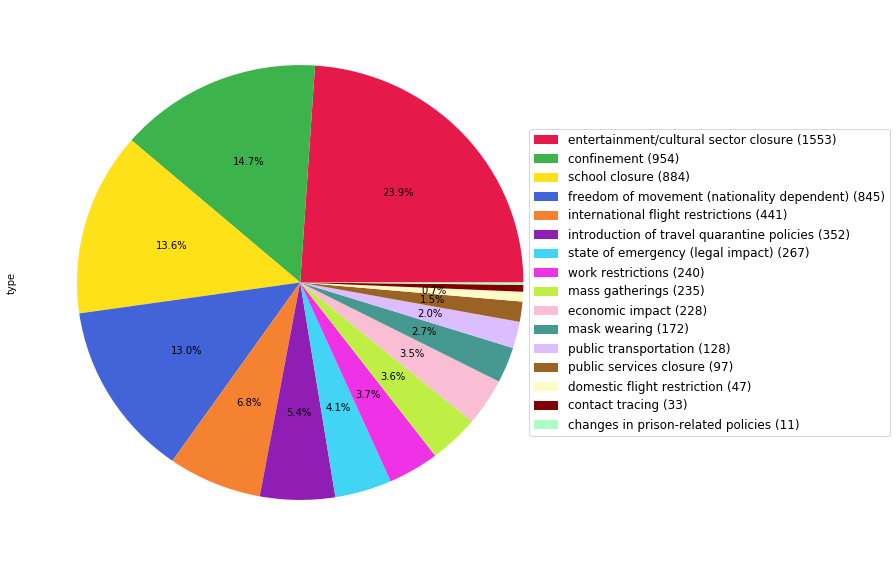}
    \caption{Distribution of \glspl{npi} in the \acrlong{wntrac} dataset.}
    \label{fig:stats-npi-distribution}
\end{figure}

\begin{figure}[htp!]
\centering
    \subcaptionbox{}{\includegraphics[height=14 \baselineskip]{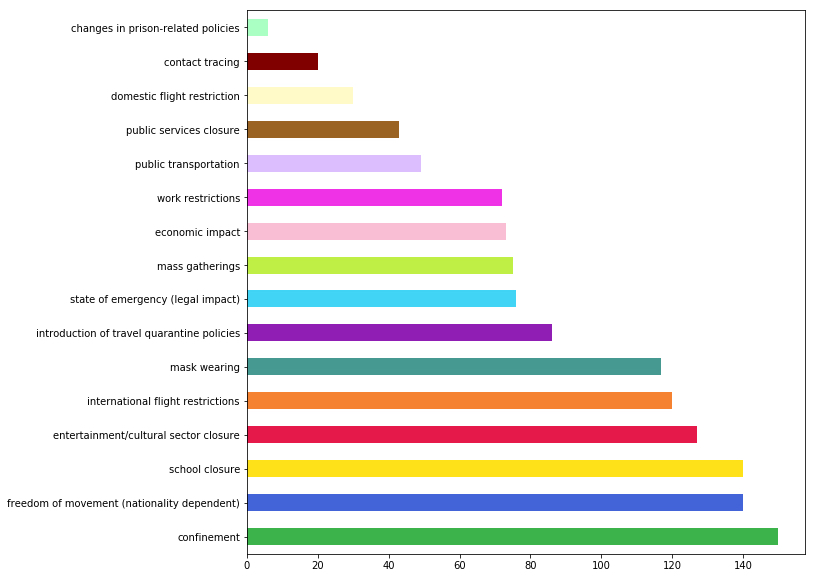}} \quad
    \subcaptionbox{}{\includegraphics[height=14 \baselineskip]{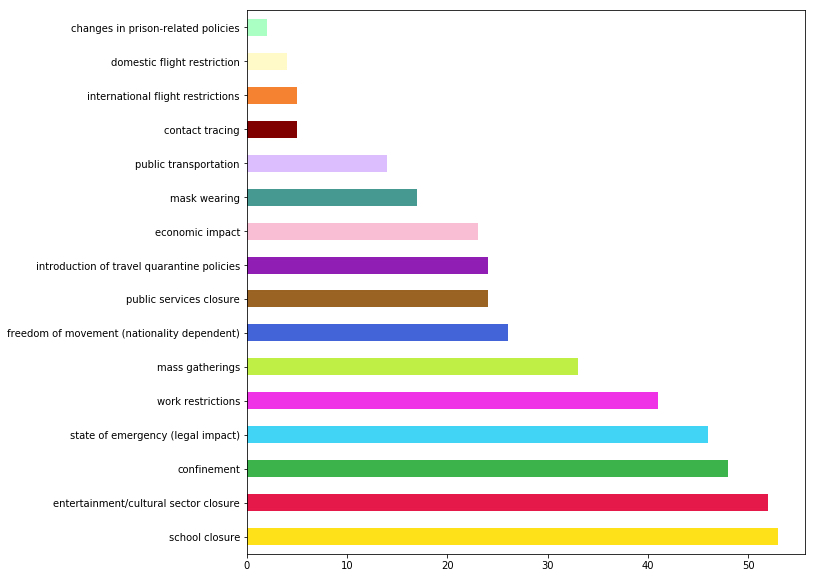}} \quad
    \caption{Number of regions implementing each \gls{npi} globally (left) and within US (right).} 
    \label{fig:stats-region-count-by-npi}
\end{figure}

\begin{figure}[htp!]
\centering
    \subcaptionbox{}{\includegraphics[height=18\baselineskip]{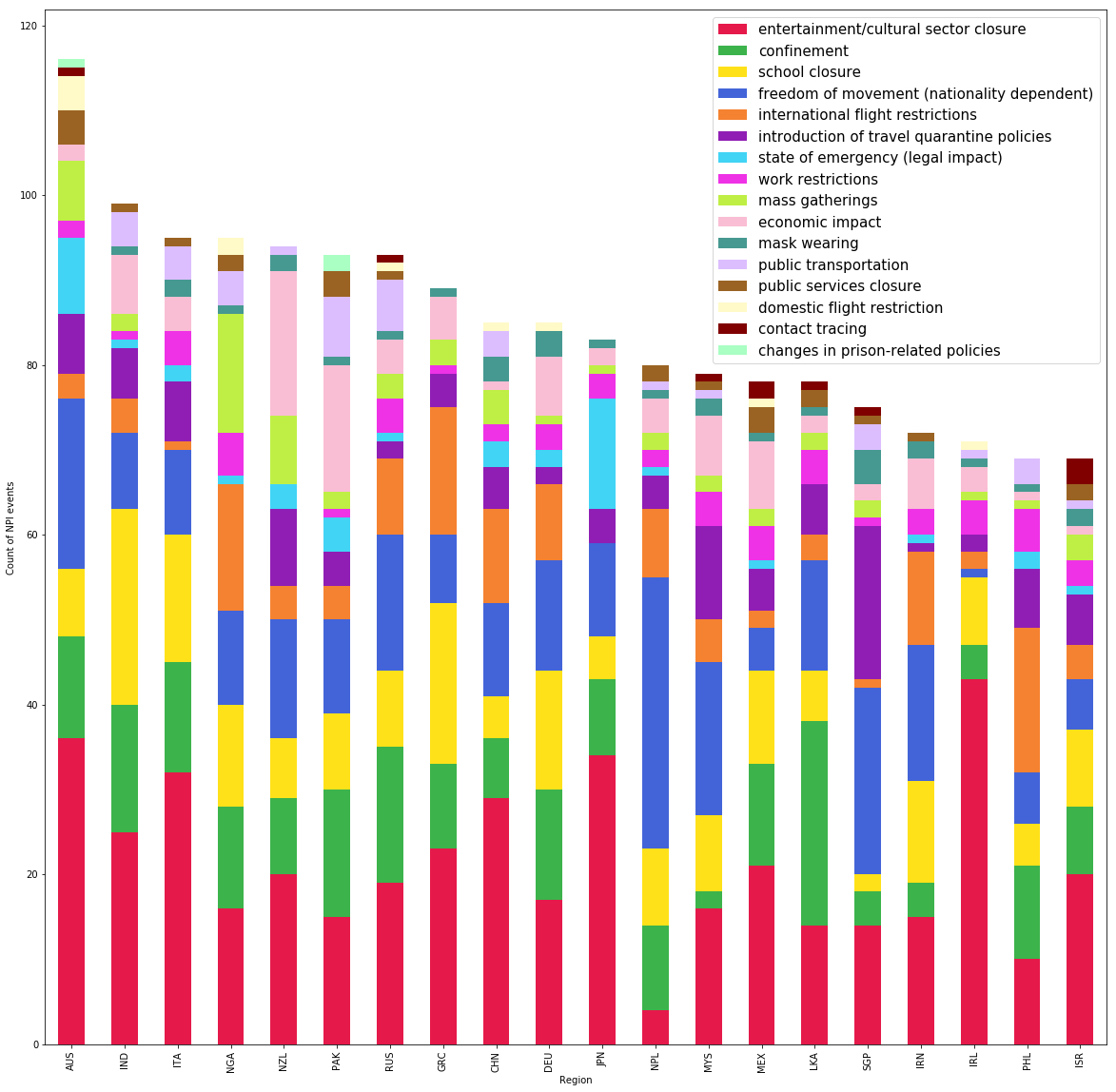}} \quad
    \subcaptionbox{}{\includegraphics[height=18\baselineskip]{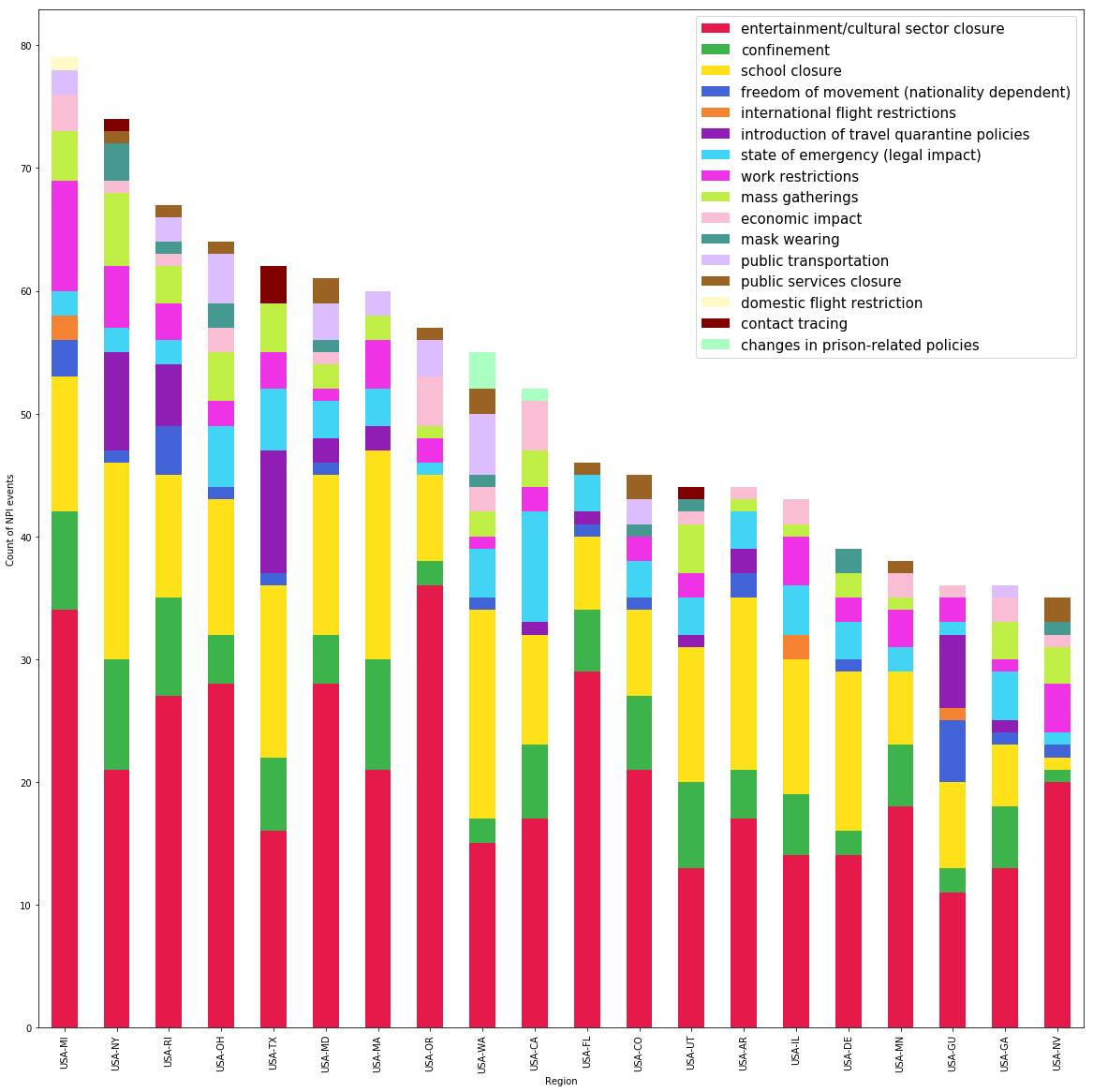}} \quad
    \caption{Distribution of \gls{npi} measures implemented in different geographies globally (left) and within US (right).} 
    \label{fig:stats-npi-count-by-region}
\end{figure}
 }

\section*{Technical Validation}\label{sec:technical-validation}

The validation team consisted of a mix of experts who participated in the design of the taxonomy and/or the pipeline and IBM volunteers who completed a brief training session about the annotation schema and tool. Validation was done in two stages. 
In the first phase, because the \gls{wntrac} tool was still being developed, we used simple CSV files to distribute the data for validation. Each annotator was given a complete \gls{document} corresponding to a Wikipedia article for a particular region, retrieved as on June 6, 2020, pre-annotated with the output of the pipeline. Each sentence was displayed in a separate line with sentences corresponding to candidate \glspl{event} highlighted with a different background color. The attributes extracted by the pipeline were listed next to each sentence. Annotators were asked to verify and correct each of these attributes. If a sentence does not discuss any of the valid event types, they were asked to mark the \gls{type} as \emph{discarded}. If a sentence was incorrectly discarded by the pipeline, they were asked to correct the \gls{type} and fill in the attributes when possible. This was, however, not uniformly enforced. In the second phase, we made \gls{tool} tool available to the annotators. The tool randomly assigns a single \gls{document} to be validated to each annotator. Each \gls{document}, consists of incremental changes to the underlying Wikipedia article since the last validation of the \gls{document}. The validation process for the second phase is similar to the first phase except that only candidate \glspl{event}, as determined by the pipeline were shown to the annotators. This time-saving move was based on the observation during the first phase, when all sentences were presented, human annotators generally agreed with the automated pipeline on discarded sentences. The \gls{nlp} model used a recall-oriented threshold and only discarded sentences with low scores on all valid \gls{npi} types. 

\iftoggle{inplace}{
 \begin{table*}[!htb]
    \centering
    \begin{tabular}{lcccccc}
        \toprule
        \multirow{2}{*}{} & \multicolumn{3}{c}{\textbf{All \gls{npi} event types}} & \multicolumn{3}{c}{\textbf{Top 5 \gls{npi} event types}} \\ \cmidrule{2-7} 
         & \textbf{A vs E\textsubscript{1}} & \textbf{A vs E\textsubscript{2}} & \textbf{E\textsubscript{1} vs E\textsubscript{2}} & \textbf{A vs E\textsubscript{1}} & \textbf{A vs E\textsubscript{2}} & \textbf{E\textsubscript{1} vs E\textsubscript{2}} \\
        \midrule
        \textbf{Type} & 0.63 & 0.69 & 0.80 & 0.81 & 0.77 & 0.85 \\ 
        \textbf{Type + Value} & 0.41 & 0.42 & 0.69 & 0.51 & 0.47 & 0.70 \\ 
        \textbf{Date} & 0.50 & 0.61 & 0.73 & 0.60 & 0.69 & 0.76 \\ 
        \textbf{Region} & 0.99 & 1.00 & 0.99 & 0.98 & 1.00 & 0.98 \\ 
        \textbf{Restriction} & 0.36 & 0.43 & 0.74 & 0.74 & 0.58 & 0.69 \\
        \textbf{Type + Date} & 0.44 & 0.53 & 0.70 & 0.51 & 0.59 & 0.72 \\ 
        \textbf{Type + Value + Date} & 0.31 & 0.34 & 0.62 & 0.35 & 0.36 & 0.59 \\ 
        \textbf{Type + Value + Date + Region} & 0.30 & 0.33 & 0.62 & 0.35 & 0.36 & 0.59 \\ 
        \textbf{Type + Value + Date + Region + Restriction} & 0.26 & 0.29 & 0.61 & 0.34 & 0.35 & 0.59 \\ \bottomrule
    \end{tabular}
    \caption{Inter-annotator agreement between average volunteers (A) and two groups of experienced volunteers (E\textsubscript{1} and E\textsubscript{2}). Region includes both country and state/territories as applicable.}
    \label{tab:iaa}
\end{table*} }

To determine the quality of the dataset post validation, \gls{iaa} was calculated on a subset, randomly sampled (2\%), from the full set that was validated by IBM volunteers. Each instance in the subset was further double annotated by two experts (randomly selected from a pool of six experts) independently, resulting in three sets of annotations per instance. The \gls{iaa} was evaluated on all five fields of the 5-tuple that uniquely defines an \gls{event}. Furthermore, the evaluation was performed at a field level for all fields except the \gls{value}, which is technically a sub-field of \gls{type} and it does not make sense to be analyzed on its own. The \gls{iaa} results are shown in Table~\ref{tab:iaa}. Note that the \gls{iaa} between experts were consistently high in all categories, indicating that the annotation schema is not ambiguous and most sentences can be consistently assigned to one of the \gls{npi} \gls{type} defined in the taxonomy. The \gls{iaa} between the volunteers and experts were also good (0.58) at the \gls{npi} \gls{type} level and the agreement is high (0.81) in the five most frequent \gls{npi} types. We plan to expand the taxonomy over time to cover more \gls{npi} types. We also plan to improve the accuracy of the pipeline by using end-to-end entity linking techniques for entity normalization and state-of-the-art methods for better temporal alignment. We plan to expand to other data sources to improve coverage. 
\section*{Usage Notes}\label{sec:usage-notes}

One of the primary objectives in creating the \gls{wntrac} dataset was to understand what types of \glspl{npi} are being implemented worldwide and to facilitate analysis of the efficacy of the different types of \glspl{npi}. Specifically, the dataset supports a variety of studies, such as correlation and analysis to understand the associations between \glspl{npi} and outcomes, causal inference between \glspl{npi} and specific outcome variables, as well as impact analysis to understand the impact on socio-economic factors. Furthermore, this dataset offers an opportunity to perform local contextualized What-if scenarios and optimal intervention planning, by incorporating  \glspl{npi} into epidemiological models. Such capabilities are critical for target decision-making to control the spread of the disease and minimize impact on society.

There are a number of questions, ranging in complexity, that the dataset can be used to answer. For example, consider the question: \emph{How many \glspl{npi} were imposed and lifted globally as the pandemic continues?}. Figure~\ref{fig:number-imposed-lifted} sums the number of \glspl{npi} imposed and lifted in all geographies per month. As expected the vast amount of \glspl{npi} were imposed during the first outbreak of \gls{covid} in March, and lifted mainly in April and May. This figure also reveals the imbalance between imposed and lifted \glspl{npi} that exists in the data. For example, while more than three thousand \glspl{npi} were imposed at March, less than five hundred were lifted between April and September. The imbalance can be the outcome of many factors, such as, how and when lifting of \glspl{npi} is announced over time. Such factors should be taken into account performing analysis using this dataset. 
\iftoggle{inplace}{
    \begin{figure}[htp!]
    \centering
    \subcaptionbox{Imposed\label{fig:imposed_count_by_month}}{\includegraphics[height=12\baselineskip,width=.45\linewidth]{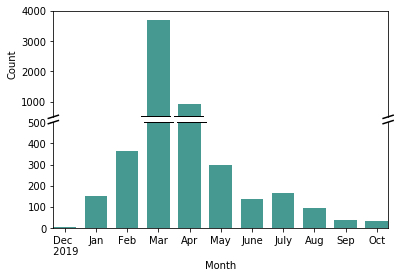}} \quad
    \subcaptionbox{Lifted\label{fig:lifted_count_by_month}}{\includegraphics[height=12\baselineskip,width=.45\linewidth]{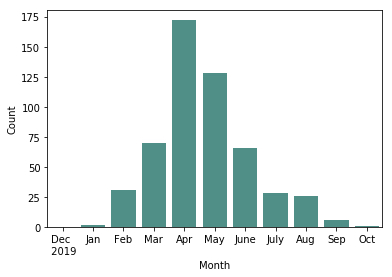}} \quad
    \caption{Number of imposed and lifted \glspl{npi} measures per month.} 
    \label{fig:number-imposed-lifted}
\end{figure}
 }
{}

A second example use of the dataset is to explore which \glspl{npi} were imposed by different countries early in the pandemic, to contain the spread of \gls{covid}?. One approach is to break the set of \glspl{npi} into two sets: travel-related and community related. Travel-related \glspl{npi} include \emph{domestic flight restrictions}, \emph{international flight restrictions}, \emph{freedom of movement (nationality dependent)}, and \emph{introduction of travel quarantine policies}. Figure~\ref{fig:travel-related} visualizes the elapsed time between the implementation of a travel-related \glspl{npi} and the recording of at least 50 cases, and time to the first reported death.  The visualization shows 9 selected regions each of which had at least one travel-related \gls{npi} among the first set of \glspl{npi} imposed in the country, and was generated by combining \gls{wntrac} dataset with \gls{covid} outcomes dataset from the World Health Organization (WHO)~\cite{who}. For each region, the blue bar plot illustrates the number of days before 50 cumulative cases, and the red points shows the number of days before the first death. From the graph, it can be observed that Singapore first imposed a travel-related \gls{npi} more than 50 days before their first death, showing an earlier response than Brazil and New York State where the first travel related \gls{npi} were imposed about 10 days after the first death. Similarly, Figure~\ref{fig:community-related} visualizes the elapsed time between the implementation of community-related \glspl{npi} and the recording of at least 50 cases and at least one death for 9 selected regions. The community-related \glspl{npi} include \emph{entertainment/cultural sector closure},  \emph{confinement}, \emph{school closure}, \emph{mass gatherings}, \emph{mask wearing}, \emph{public services closure}, \emph{public transportation}, \emph{work restrictions}, and \emph{state of emergency}. It can be noted that at least one community-related \gls{npi} was imposed for each of the selected regions prior to their first recorded death due to COVID-19. 
\iftoggle{inplace}{
    \begin{figure}[htp!]
    \centering
    \subcaptionbox{Travel-related \glspl{npi}\label{fig:travel-related}}{\includegraphics[height=12\baselineskip,width=.45\linewidth]{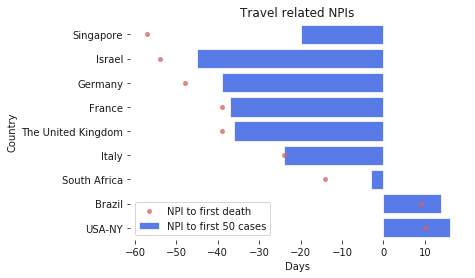}} \quad
    \subcaptionbox{Community-related \glspl{npi}\label{fig:community-related}}{\includegraphics[height=12\baselineskip,width=.45\linewidth]{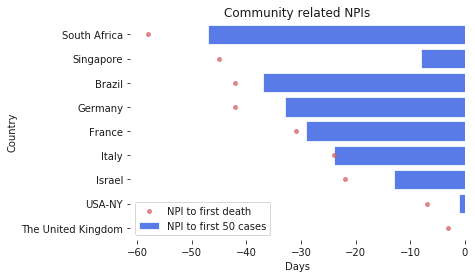}} \quad
    \caption{Elapsed time (in days) between the introduction of \glspl{npi} and recording of first death (\textcolor{red}{red}) or 50 cases (\textcolor{blue}{blue}) in countries that implemented travel-related vs community-related \glspl{npi} first.} 
    \label{fig:travel-community-related}
\end{figure}
 }
\\

As a third example, we demonstrate how the \gls{wntrac} dataset can be used to generate an index, a summary statistic between $[0,1]$ that represents the NPIs imposed and, if available, the adherence. This index can be used to study the relationship between \glspl{npi} and \gls{covid} outcomes over time and to compare response strategies across jurisdictions. Figure~\ref{fig:npi-trends} illustrates this using data from representative states in the United States (Florida, Georgia, New York, and Texas). In the figure, the bar graph shows the trend for the exponentially weighted moving average of new cases per 100,000 population. The red continuous line is the proportion of the \gls{npi} (out of thirteen \gls{npi} \glspl{type} in the \gls{wntrac} dataset) that a region has imposed at a given time. The blue continuous line is the \gls{wntrac} \gls{npi-index}, a composite index that captures both the stringency levels of the \glspl{npi} and community mobility data as a proxy measure of adherence to \glspl{npi} strategies. The \gls{wntrac} \gls{npi-index}, denoted $\eta(t)$,  is presented in Eq. \ref{eq:1}, and the code for the \gls{wntrac} \gls{npi-index} is  available in the repository.

\begin{equation} \label{eq:1}
  \eta(t) = \omega_{0}SI(t) + \omega_{1}\frac{e^{A(t)}}{1 + e^{A(t)}},
\end{equation}

$\omega_{0}, \omega_{1} > 0$ are weights applied to each term and $\omega_{0} + \omega{1} = 1$. Specifically, the first term, $SI$, is derived from mapping and scoring the \gls{wntrac} \gls{npi} similarly as presented in the \gls{oxcgrt} stringency index~\cite{hale2020oxford}. The second term represents adherence at a specific point in time, $A(t)$, by using mobility data as a proxy. Specifically, we define $A(t)$ in Eq.~\ref{eq:2} as a function of the "anticipated mobility", $m_{ant}$, and the "observed mobility," $m_{obs}$. The anticipated mobility at a specific point in time is the mobility score that would potentially be associated with the \glspl{npi} at that time. The observed mobility is the mobility value observed in that region at a specific time point and ideally should be close the value of anticipated mobility. In our work, we assume a negative relations between stringency and mobility, and anticipated mobility is derived from this linear relationship with noise.

\begin{equation} \label{eq:2}
  A(t) = \frac{m_{ant} - m_{obs}}{m_{ant}}.
\end{equation}

As illustrated, the \gls{wntrac} \gls{npi} metrics can be compared to existing metrics such as the \gls{oxcgrt} stringency index\cite{hale2020oxford}. Of note is the detailed interpretation of the relationships illustrated in this example is subject to addressing limitations such as missing data and will be pursued as part of our future work.

\iftoggle{inplace}{
    
\begin{figure}[htp!]
    \captionsetup[subfigure]{labelformat=empty}
    \centering
    \subcaptionbox{}{\includegraphics[height=10\baselineskip,width=.4\linewidth]{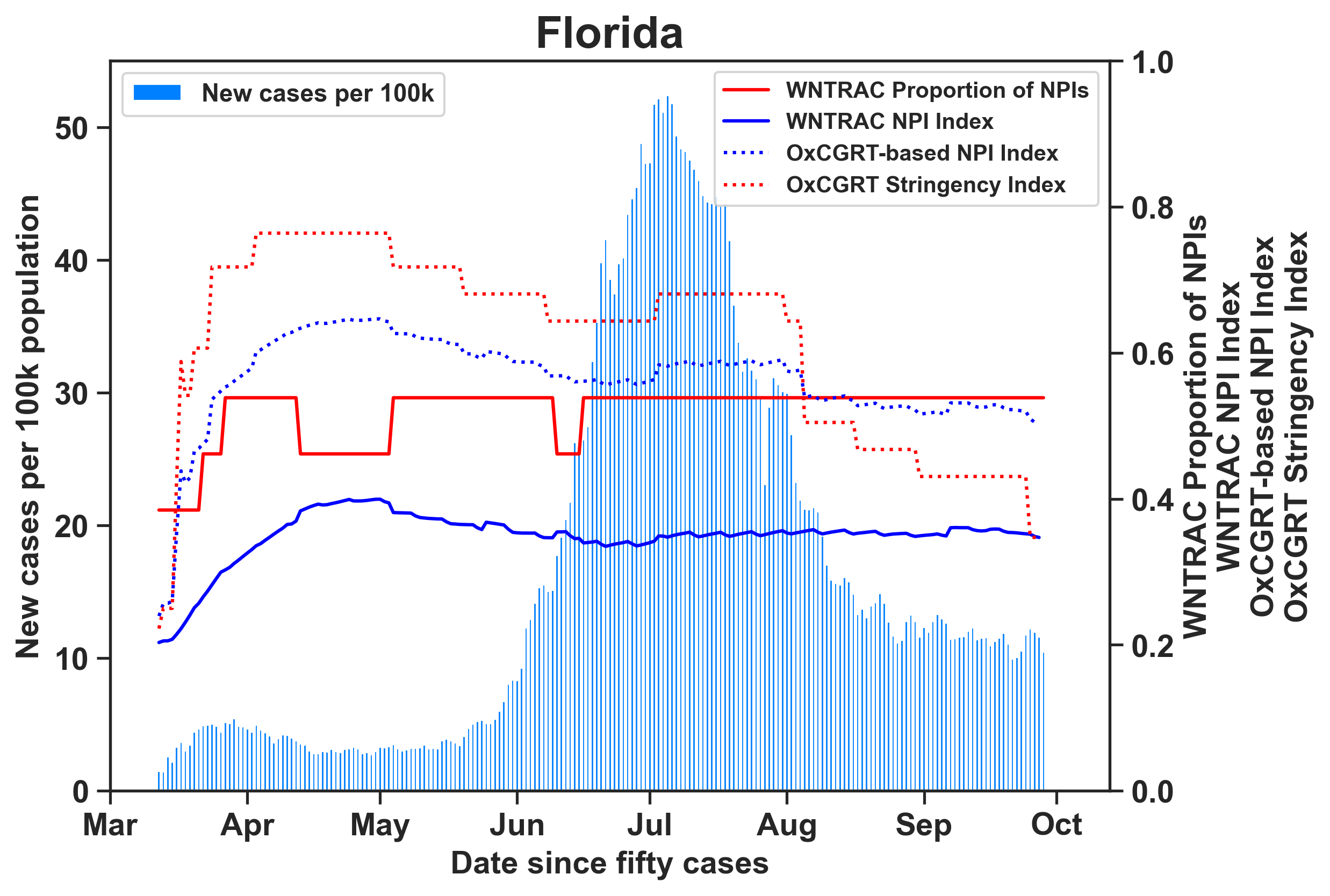}}\quad
    \subcaptionbox{}{\includegraphics[height=10\baselineskip,width=.4\linewidth]{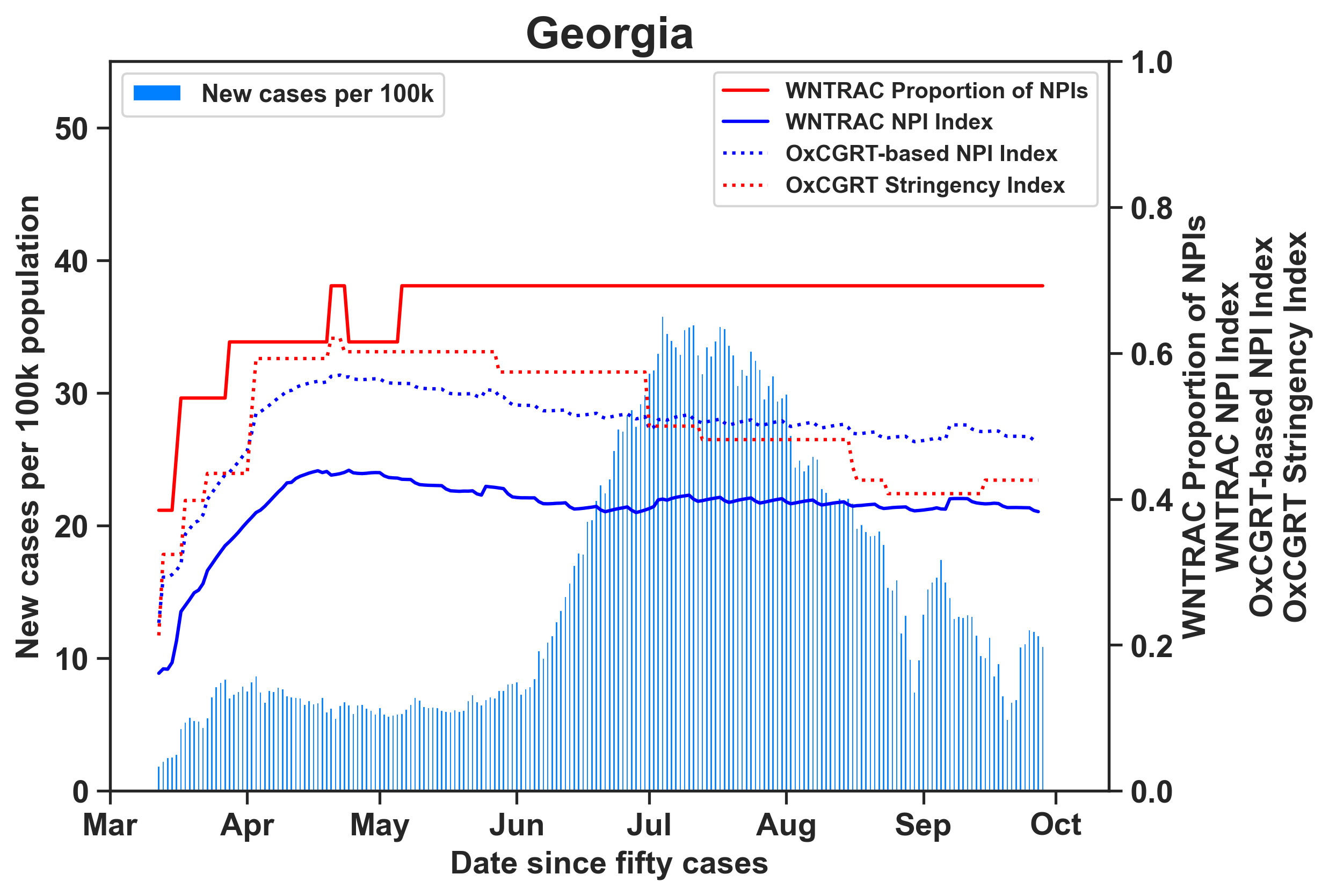}} \quad
    \subcaptionbox{}{\includegraphics[height=10\baselineskip,width=.4\linewidth]{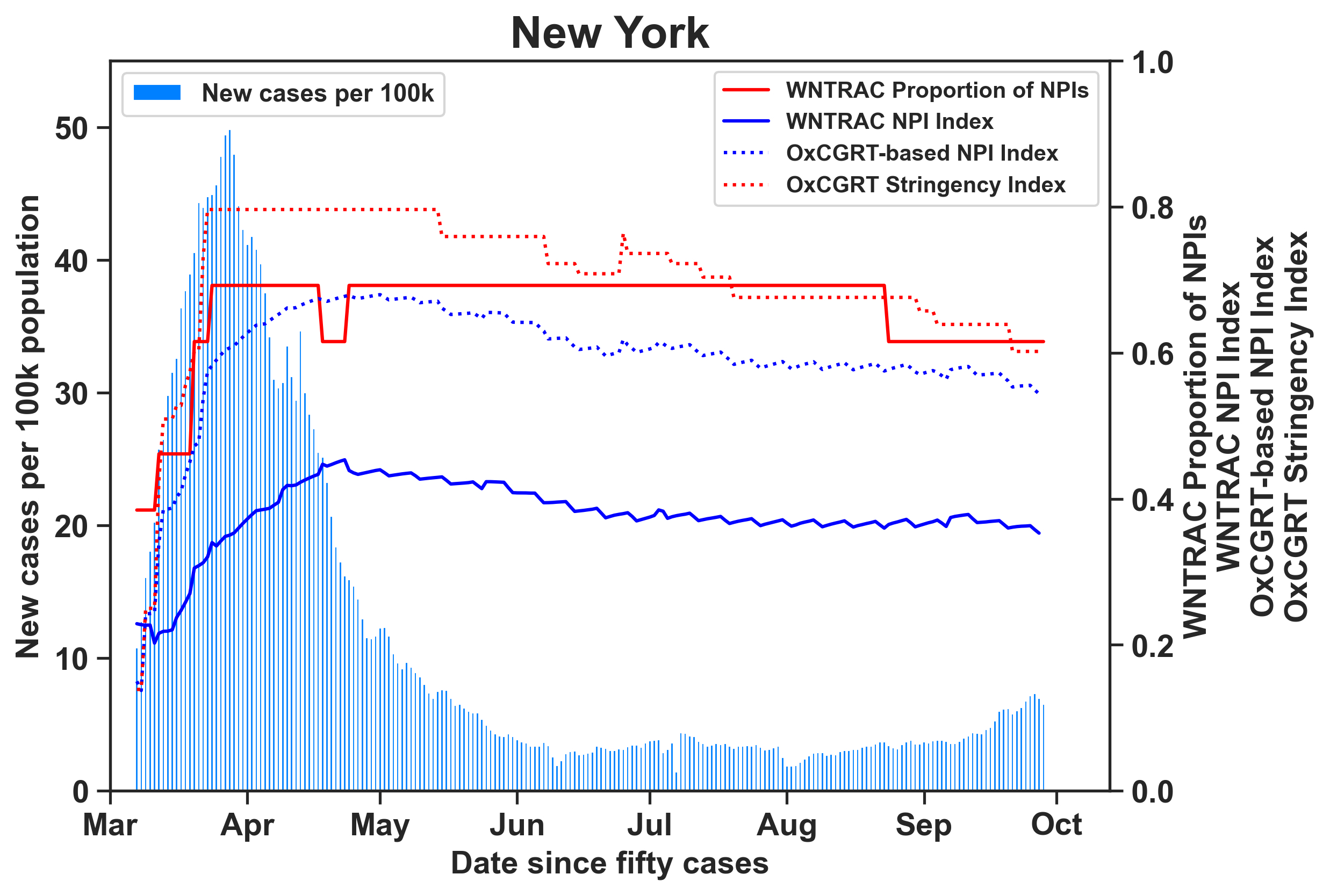}} \quad
    \subcaptionbox{}{\includegraphics[height=10\baselineskip,width=.4\linewidth]{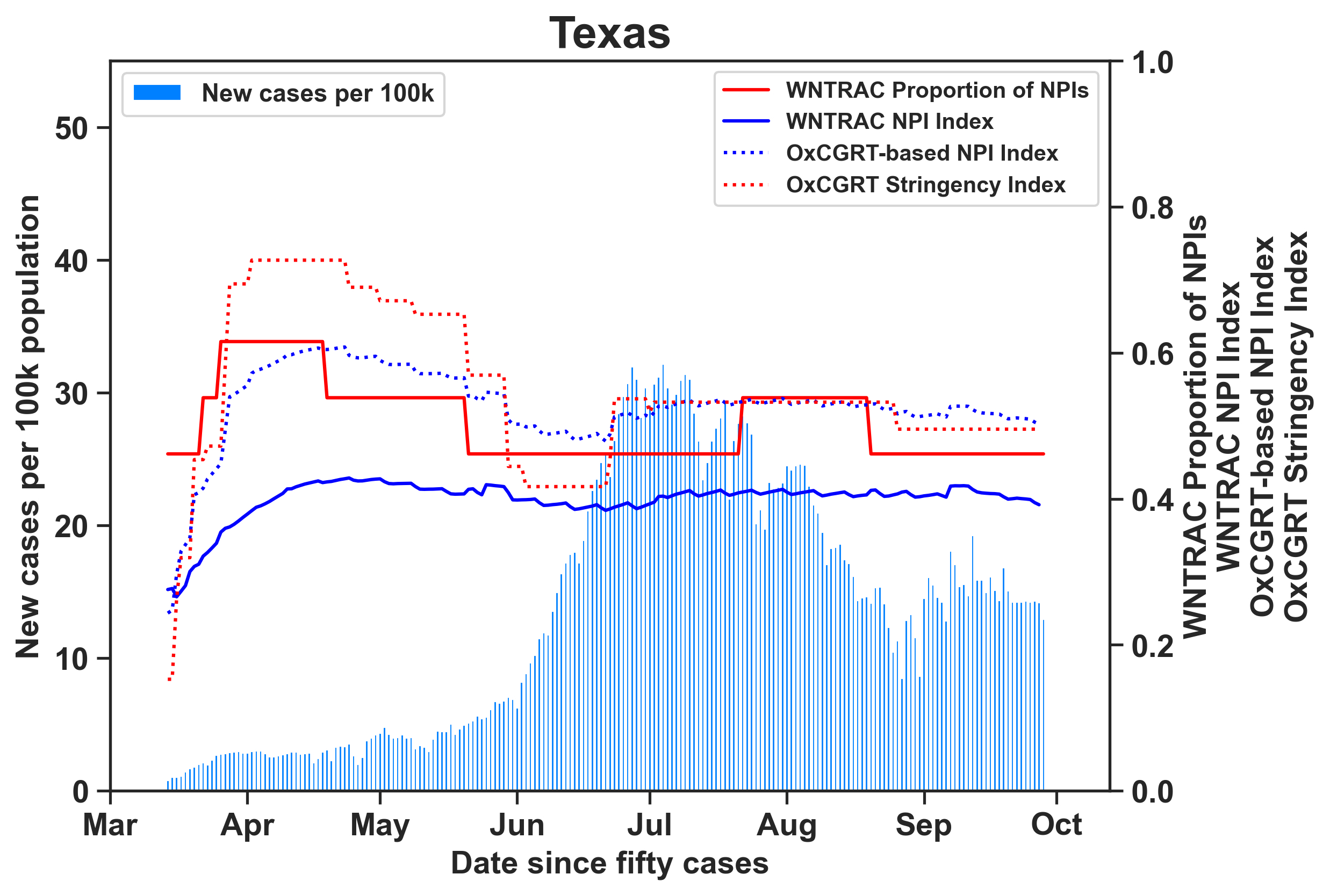}} \quad
    \caption{Trends in \gls{covid} cases per 100,1000 population and the \gls{npi}-based indices in representative US states.} 
    \label{fig:npi-trends}
\end{figure}

 }

Finally, another important application of the \gls{wntrac} dataset is to support What-if analysis and decision-making for optimal intervention planning. This is especially important to provide critical, time-sensitive decision support to various leaders, and decision-making teams such as \gls{covid} task force teams as they determine which \glspl{npi} to impose or lift over time. Efficiency in this decision-making process is important, as the space of all potential combinations and variations of \glspl{npi} is large and complex. The options for a particular region have varying degrees of impact on outcomes for that region. Tools ~\cite{notnets} that enable what-if analysis and intervention planning, at both national and sub-national levels, that incorporate the \gls{wntrac} dataset can be leveraged to meet this need. For decision-makers, these tools enable easy navigation through the complex intervention space in a timely manner to generate the most optimal and context-relevant  \gls{covid} intervention programs. A key requirement for such tools are epidemiological models that are calibrated in such a way that the resulting forecasts can be trusted as accurate projections. To calibrate these models, it is critical to consider the \gls{npi} that have been imposed so that the drivers of disease spread can be contextualized for a region. By incorporating \gls{npi} into the models improved projections of outcomes of the disease can be generated, yielding more accurate scenarios for decision-makers to explore.

In addition to the above examples, the \gls{wntrac} dataset can be used to support other objectives, including estimating the relationships between \glspl{npi} and
\begin{itemize} [noitemsep, topsep=0pt]
    \item consumers behavior by, for example, correlating between retail data and \glspl{npi}.
    \item environmental changes such as pollution levels. 
     \item actual compliance by the population. Naturally, not all the interventions recorded in the dataset are an accurate representation of reality as some of the interventions capture a governmental request that might not be followed by the entire population. Thus, it might be useful to integrate the \gls{wntrac} dataset with other publicly available data sources that can provide information regarding the level of compliance with an intervention, such as mobility information~\cite{apple-mobility, google-mobility}, where we provided an example with the NPI-Index above, and social media.
\end{itemize}
Lastly, one other interesting use case is to estimate the economic impact of \glspl{npi} by, for example, relating unemployment rates and jurisdictional debt with \glspl{npi}. Estimation of the effect of \glspl{npi} on non-\gls{covid} health problems, such as late cancer detection due to missed screening tests, will also be useful.

\iffalse Procedurally, this involves a series of preprocessing steps that include loading the \gls{wntrac} dataset, selecting the subset of 13 \glspl{npi} categories used for subsequent analyses, identifying country-level \gls{npi} events for each of the selected \gls{npi} categories, and identifying the date when each of the selected \gls{npi} categories was first imposed in each country. Subsequently, we combined the preprocessed \gls{npi} data with \gls{covid} outcomes data from the World Health Organization (WHO) \cite{who}. Note that the WHO dataset contains daily counts of new cases, cumulative cases, new deaths, and cumulative deaths reported per country/territory. In our example analysis, for each country/territory in the preprocessed NPI data, we used WHO data to identify the dates when the country/territory first reported 50 cumulative cases and first deaths, and excluded countries/territories with fewer than 50 cases, no deaths, or no data. Consequently, the final combined dataset used for our example analyses was in a longitudinal format where each row represents daily counts of outcomes per territory from the WHO, along with binary indicator variables for each \gls{npi} derived from \gls{wntrac} that denote whether or not the \gls{npi} was imposed or not imposed.
\fi

\section*{Code Availability}\label{sec:code-availability}
The source code for the \gls{wntrac} automated \gls{npi} curation system, including the data processing pipeline, \gls{tool} tool and \gls{npi} data browser is available in a public GitHub repository at~\url{https://github.com/IBM/wntrac/tree/master/code} along side the up-to-date version of the dataset~\url{https://github.com/IBM/wntrac/tree/master/data}. Please refer to the README file in repository for further instructions on using the code.

\section*{Acknowledgements}\label{sec:ack} 
We thank IBM Research volunteers for validation and maintenance of the WNTRAC dataset. 

\section*{Author Contributions Statement}\label{sec:contribution}
IBM Research Haifa team identified the need for the dataset, defined the taxonomy of \glspl{npi} based on requirements for epidemiological modeling and developed the validation guidelines for volunteers. IBM Research Yorktown Heights team developed \gls{nlp} for \gls{npi} extraction, developed the semi-automated system to construct the dataset and keep it current and built the \gls{tool} tool. IBM Research Nairobi team designed and implemented graphical user interface for the \gls{npi} data browser for end users to browse, query and visualize the dataset and the associated descriptive statistics. Senior authors Michal Rosen-Zvi, Divya Pathak and Aisha Walcott-Bryant lead the respective teams.   
\section*{Competing Interests}\label{sec:compete} 
The authors declare no competing interests.

\iftoggle{inplace}{
    \clearpage
\begin{table}[!htb]
    \begin{minipage}{.33\linewidth}
      \centering
      \scriptsize
        \begin{tabular}{ll}
        \textbf{Region} & \textbf{Code} \\ \midrule
        Abkhazia & GEO-AB \\
        Afghanistan & AFG \\
        Albania & ALB \\
        Algeria & DZA \\
        Andorra & AND \\
        Angola & AGO \\
        Anguilla & AIA \\
        Antarctica & ATA \\
        Antigua and Barbuda & ATG \\
        Argentina & ARG \\
        Armenia & ARM \\
        Australia & AUS \\
        Austria & AUT \\
        Azerbaijan & AZE \\
        Bahamas & BHS \\
        Bahrain & BHR \\
        Bangladesh & BGD \\
        Barbados & BRB \\
        Belarus & BLR \\
        Belgium & BEL \\
        Belize & BLZ \\
        Benin & BEN \\
        Bhutan & BTN \\
        Bolivia & BOL \\
        Bosnia and Herzegovina & BIH \\
        Botswana & BWA \\
        Brazil & BRA \\
        Brunei & BRN \\
        Bulgaria & BGR \\
        Burkina Faso & BFA \\
        Burundi & BDI \\
        Cambodia & KHM \\
        Cameroon & CMR \\
        Canada & CAN \\
        Cape Verde & CPV \\
        Central African Republic & CAF \\
        Chad & TCD \\
        Chile & CHL \\
        China & CHN-TB \\
        Colombia & COL \\
        Comoros & COM \\
        Costa Rica & CRI \\
        Croatia & HRV \\
        Cuba & CUB \\
        Cyprus & CYP \\
        Czech Republic & CZE \\
        Democratic Republic of Congo & COD \\
        Denmark & DNK \\
        Djibouti & DJI \\
        Dominica & DMA \\
        Dominican Republic & DOM \\
        Ecuador & ECU \\
        Egypt & EGY \\
        El Salvador & SLV \\
        Equatorial Guinea & GNQ \\
        Eritrea & ERI \\
        Estonia & EST \\
        Eswatini & SWZ \\
        Ethiopia & ETH \\
        Faroe Islands & DEN-FI \\
        Fiji & FJI \\
        Finland & FIN \\
        France & FRA \\
        French Guiana & GUF \\
        French Polynesia & PYF \\
        French Saint Martin & MAF \\
        Gabon & GAB \\
        Gambia & GMB \\
        Georgia (country) & GEO \\
        \end{tabular}
    \end{minipage}\begin{minipage}{.33\linewidth}
      \centering
      \scriptsize
        \begin{tabular}{ll}
        Germany & DEU \\
        Ghana & GHA \\
        Greece & GRC \\
        Greenland & DEN-GR \\
        Grenada & GRD \\
        Guadeloupe & FRA-GU \\
        Guatemala & GTM \\
        Guinea & GIN \\
        Guinea-Bissau & GNB \\
        Guyana & GUY \\
        Haiti & HTI \\
        Honduras & HND \\
        Hong Kong & CHN-HK \\
        Hungary & HUN \\
        Iceland & ISL \\
        India & IND \\
        Indonesia & IDN \\
        Iran & IRN \\
        Iraq & IRQ \\
        Israel & ISR \\
        Italy & ITA \\
        Ivory Coast & CIV \\
        Jamaica & JAM \\
        Japan & JPN \\
        Jordan & JOR \\
        Kazakhstan & KAZ \\
        Kenya & KEN \\
        Kosovo & KOS \\
        Kuwait & KWT \\
        Kyrgyzstan & KGZ \\
        Laos & LAO \\
        Latvia & LVA \\
        Lebanon & LBN \\
        Lesotho & LSO \\
        Liberia & LBR \\
        Libya & LBY \\
        Liechtenstein & LIE \\
        Lithuania & LTU \\
        Luhansk People's Republic & UKR-09 \\
        Luxembourg & LUX \\
        Madagascar & MDG \\
        Mainland China & CHN \\
        Malawi & MWI \\
        Malaysia & MYS \\
        Maldives & MDV \\
        Mali & MLI \\
        Malta & MLT \\
        Martinique & FRA-MA \\
        Mauritania & MRT \\
        Mauritius & MUS \\
        Mayotte & FRA-MA \\
        Mexico & MEX \\
        Moldova & MDA \\
        Monaco & MCO \\
        Mongolia & MNG \\
        Montenegro & MNE \\
        Morocco & MAR \\
        Mozambique & MOZ \\
        Myanmar & MMR \\
        Namibia & NAM \\
        Nepal & NPL \\
        Netherlands & NLD \\
        New Caledonia & FRA-NC \\
        New Zealand & NZL \\
        Nicaragua & NIC \\
        Niger & NER \\
        Nigeria & NGA \\
        North Korea & PRK \\
        North Macedonia & MKD \\
        Northern Cyprus & CYP \\
        \end{tabular}
    \end{minipage} 
    \begin{minipage}{.33\linewidth}
      \scriptsize
        \begin{tabular}{ll}
        Norway & NOR \\
        Oman & OMN \\
        Pakistan & PAK \\
        Panama & PAN \\
        Papua New Guinea & PNG \\
        Paraguay & PRY \\
        Peru & PER \\
        Philippines & PHL \\
        Poland & POL \\
        Portugal & PRT \\
        Qatar & QAT \\
        Republic of Artsakh & ARM \\
        Republic of Congo & COG \\
        Republic of Ireland & IRL \\
        Romania & ROU \\
        Russia & RUS \\
        Rwanda & RWA \\
        Saint Helena & SHN \\
        Saint Kitts and Nevis & KNA \\
        Saint Lucia & LCA \\
        Saint Pierre and Miquelon & SPM \\
        Saint Vincent and Grenadines & VCT \\
        San Marino & SMR \\
        Saudi Arabia & SAU \\
        Senegal & SEN \\
        Serbia & SRB \\
        Seychelles & SYC \\
        Sierra Leone & SLE \\
        Singapore & SGP \\
        Slovakia & SVK \\
        Slovenia & SVN \\
        Somalia & SOM \\
        Somalia & SOM \\
        South Africa & ZAF \\
        South Korea & KOR \\
        South Sudan & SSD \\
        Spain & ESP \\
        Sri Lanka & LKA \\
        State of Palestine & PSE \\
        Sudan & SDN \\
        Suriname & SUR \\
        Sweden & SWE \\
        Switzerland & CHE \\
        Syria & SYR \\
        Taiwan & TWN \\
        Tajikistan & TJK \\
        Tanzania & TZA \\
        Thailand & THA \\
        Timor-Leste & TLS \\
        Togo & TGO \\
        Trinidad and Tobago & TTO \\
        Tunisia & TUN \\
        Turkey & TUR \\
        Turkmenistan & TKM \\
        Uganda & UGA \\
        United Arab Emirates & ARE \\
        United Kingdom & GBR \\
        United States & USA \\
        Uruguay & URY \\
        Uzbekistan & UZB \\
        Vatican City & VAT \\
        Venezuela & VEN \\
        Vietnam & VNM \\
        Western Sahara & ESH \\
        Yemen & YEM \\
        Zambia & ZMB \\
        Zimbabwe & ZWE \\
                 & \\
                 & \\
                 & \\
        \end{tabular}
    \end{minipage} 
    \caption{List of regions currently supported by the \gls{wntrac} dataset.}
    \label{tab:regions-global}
\end{table}

\clearpage

\begin{table}[!htb]
    \begin{minipage}{.5\linewidth}
      \centering
      \scriptsize
        \begin{tabular}{ll}
        \textbf{Region} & \textbf{Code} \\ \midrule
        Alabama & USA-AL \\
        Alaska & USA-AK \\
        Arizona & USA-AZ \\
        Arkansas & USA-AR \\
        California & USA-CA \\
        Colorado & USA-CO \\
        Connecticut & USA-CT \\
        Delaware & USA-DE \\
        Florida & USA-FL \\
        Georgia (U.S. state) & USA-GA \\
        Guam & USA-GU \\
        Guantanamo Bay Naval Base & USA-Guantanamo\_Bay\_Naval\_Base \\
        Hawaii & USA-HI \\
        Idaho & USA-ID \\
        Illinois & USA-IL \\
        Indiana & USA-IN \\
        Iowa & USA-IA \\
        Kansas & USA-KS \\
        Kentucky & USA-KY \\
        Louisiana & USA-LA \\
        Maine & USA-ME \\
        Maryland & USA-MD \\
        Massachusetts & USA-MA \\
        Michigan & USA-MI \\
        Minnesota & USA-MN \\
        Mississippi & USA-MS \\
        Missouri & USA-MO \\
        Montana & USA-MT \\
        \end{tabular}
    \end{minipage}\begin{minipage}{.5\linewidth}
      \centering
      \scriptsize
        \begin{tabular}{ll}
        Nevada & USA-NV \\
        New Hampshire & USA-NH \\
        New Jersey & USA-NJ \\
        New Mexico & USA-NM \\
        New York (state) & USA-NY \\
        North Carolina & USA-NC \\
        North Dakota & USA-ND \\
        Northern Mariana Islands & USA-Northern\_Mariana\_Islands \\
        Ohio & USA-OH \\
        Oklahoma & USA-OK \\
        Oregon & USA-OR \\
        Pennsylvania & USA-PA \\
        Puerto Rico & USA-PR \\
        Rhode Island & USA-RI \\
        South Carolina & USA-SC \\
        South Dakota & USA-SD \\
        Tennessee & USA-TN \\
        Texas & USA-TX \\
        United States Virgin Islands & USA-Virgin\_Islands \\
        Utah & USA-UT \\
        Vermont & USA-VT \\
        Virginia & USA-VA \\
        Washington (state) & USA-WA \\
        Washington D.C. & USA-DC \\
        West Virginia & USA-WV \\
        Wisconsin & USA-WI \\
        Wyoming & USA-WY \\
        \end{tabular}
    \end{minipage} 
    \caption{List of US states and territories currently supported by the \gls{wntrac} dataset.}
    \label{tab:regions-us}
\end{table}  }

\nottoggle{inplace}{
    \clearpage
    \section*{Figures and Tables}\label{sec:figures-tables} 
    \begin{figure}[htp!]
  \centering
  \includegraphics[width=0.8\linewidth]{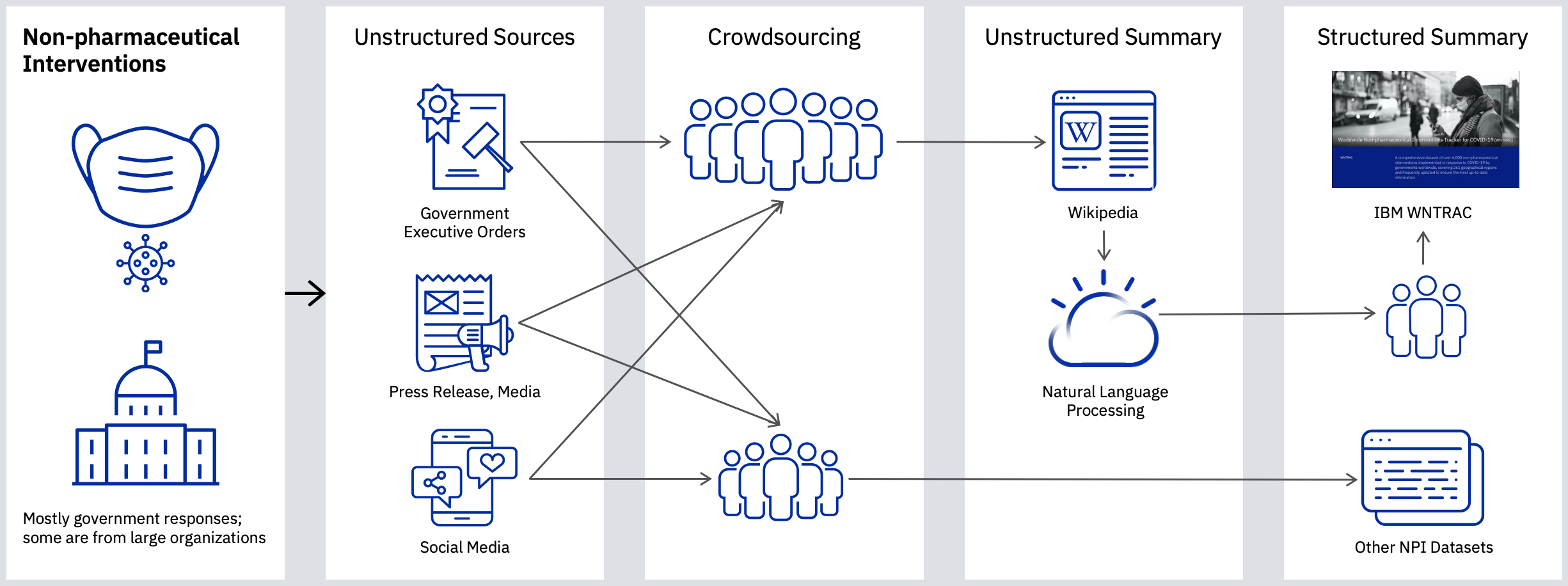}
  \caption{Artificial intelligence assisted approach to build the \gls{wntrac} dataset.}  
  \label{fig:approach}
\end{figure} \begin{figure}[htp!]
  \centering
  \includegraphics[width=0.6\linewidth]{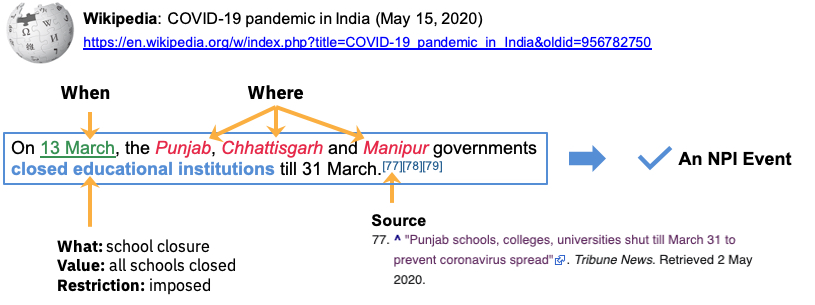}
  \caption{An example of the \gls{npi} measure mentioned in the Wikipedia article of 15\textsuperscript{th} May 2020.}  
  \label{fig:npi-example-may-15}
\end{figure}

\iffalse
    \begin{figure}[htp!]
      \centering
      \includegraphics[width=0.6\linewidth]{wntrac-eg-june-5}
      \caption{An example of the \gls{npi} event reported in the Wikipedia article on 5\textsuperscript{th} June 2020. Compared to the May 15 version of Figure~\ref{fig:npi-example-may-15}, the regions and sources have been updated. }  
      \label{fig:npi-example-june-5}
    \end{figure}
\fi \begin{figure}[htp!]
  \centering
  \includegraphics[width=0.6\linewidth]{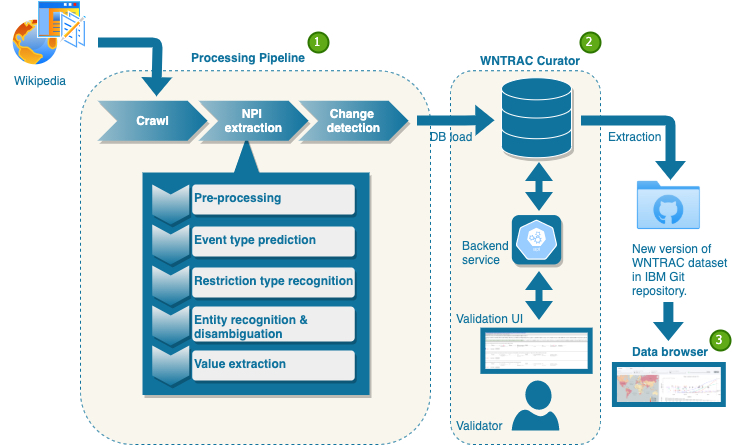}
  \caption{The \gls{wntrac} automated \gls{npi} curation system. It consists of a processing pipeline, \gls{tool} validation tool, and \gls{npi} data browser.}  
  \label{fig:system}
\end{figure} \begin{figure}[htp!]
    \centering
    \includegraphics[width=0.99\linewidth]{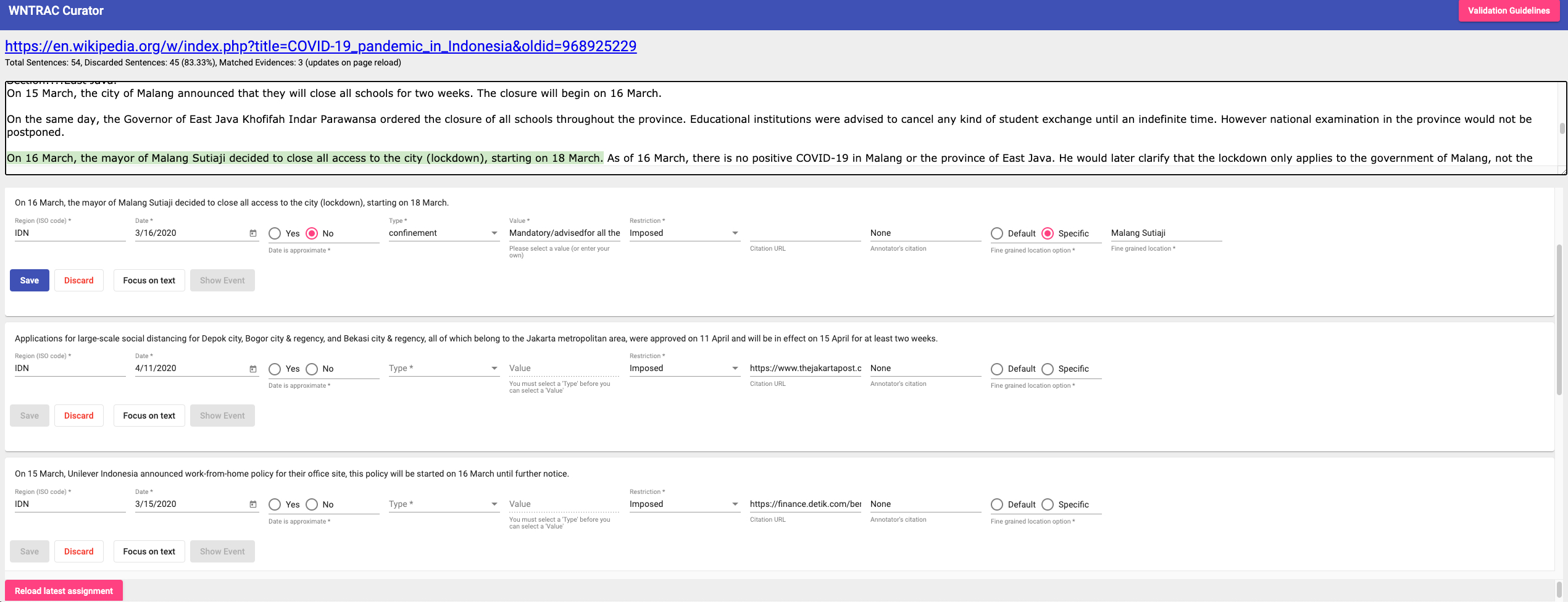}
    \caption{\gls{tool} tool used for ongoing validation of the dataset.}
    \label{fig:wntrac-curator}
\end{figure} \begin{figure}[htp!]
    \centering
    \includegraphics[width=0.9\linewidth]{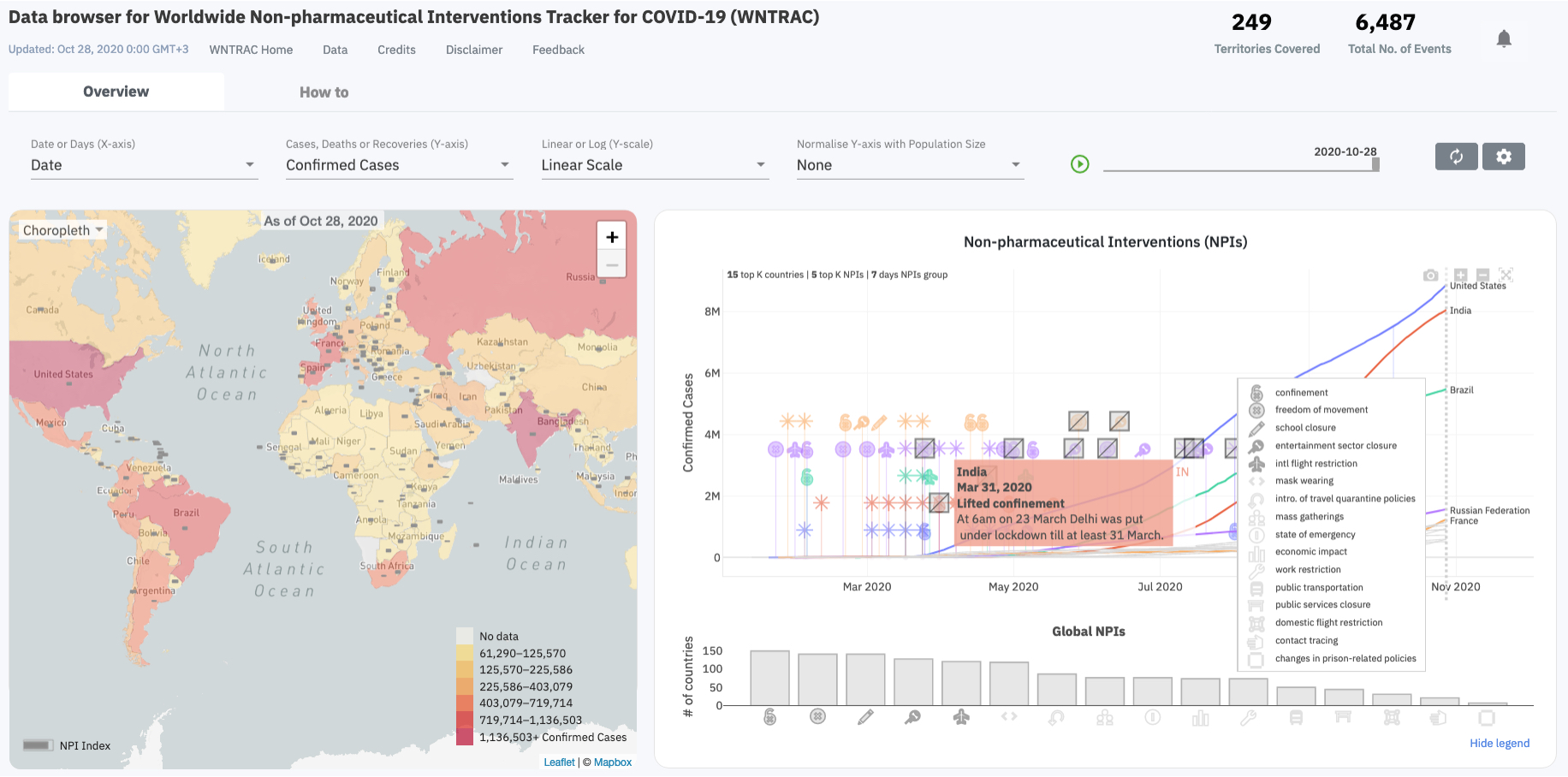}
    \caption{Data browser for visualizing the \acrlong{wntrac} dataset.}
    \label{fig:data_browser}
\end{figure} \begin{figure}[htp!]
\centering
    \includegraphics[width=0.5\linewidth]{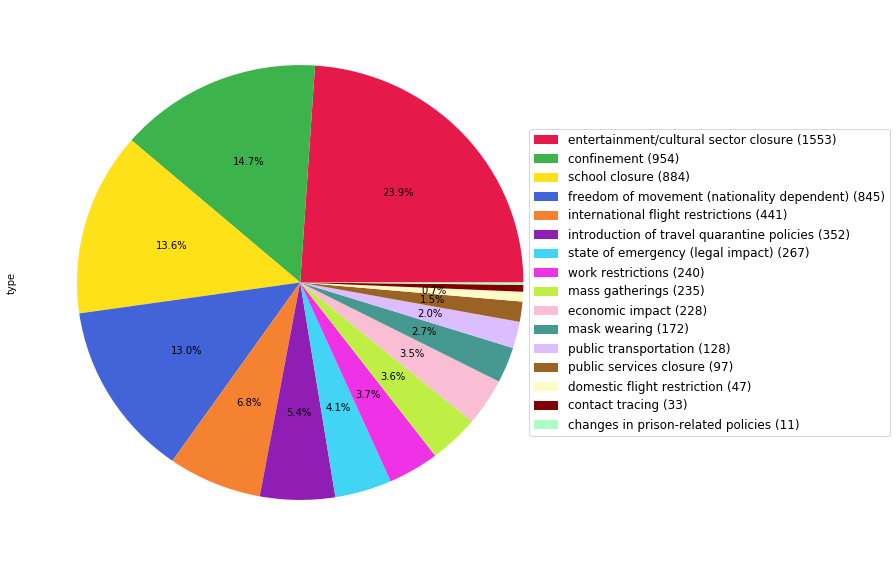}
    \caption{Distribution of \glspl{npi} in the \acrlong{wntrac} dataset.}
    \label{fig:stats-npi-distribution}
\end{figure}

\begin{figure}[htp!]
\centering
    \subcaptionbox{}{\includegraphics[height=14 \baselineskip]{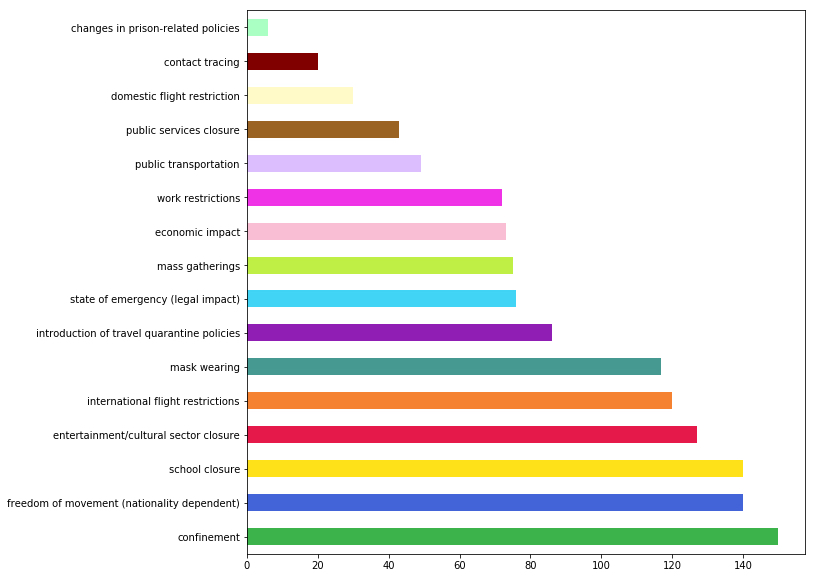}} \quad
    \subcaptionbox{}{\includegraphics[height=14 \baselineskip]{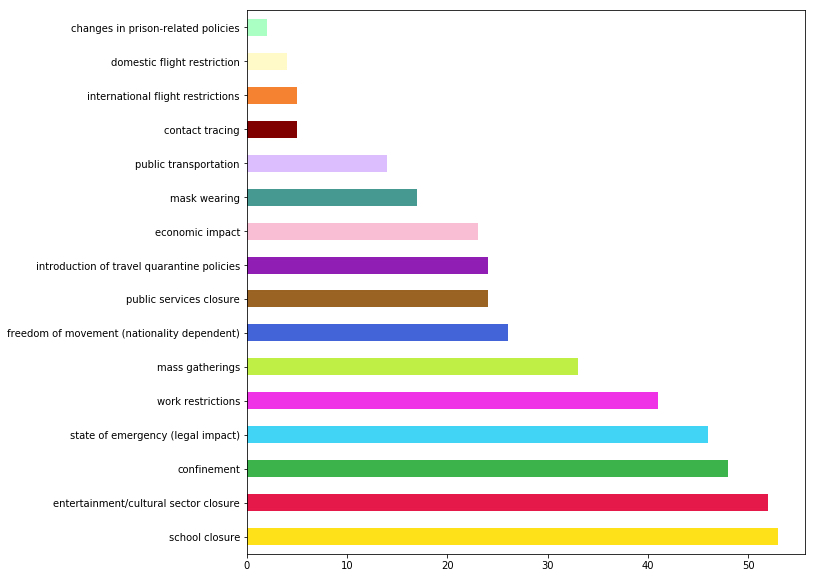}} \quad
    \caption{Number of regions implementing each \gls{npi} globally (left) and within US (right).} 
    \label{fig:stats-region-count-by-npi}
\end{figure}

\begin{figure}[htp!]
\centering
    \subcaptionbox{}{\includegraphics[height=18\baselineskip]{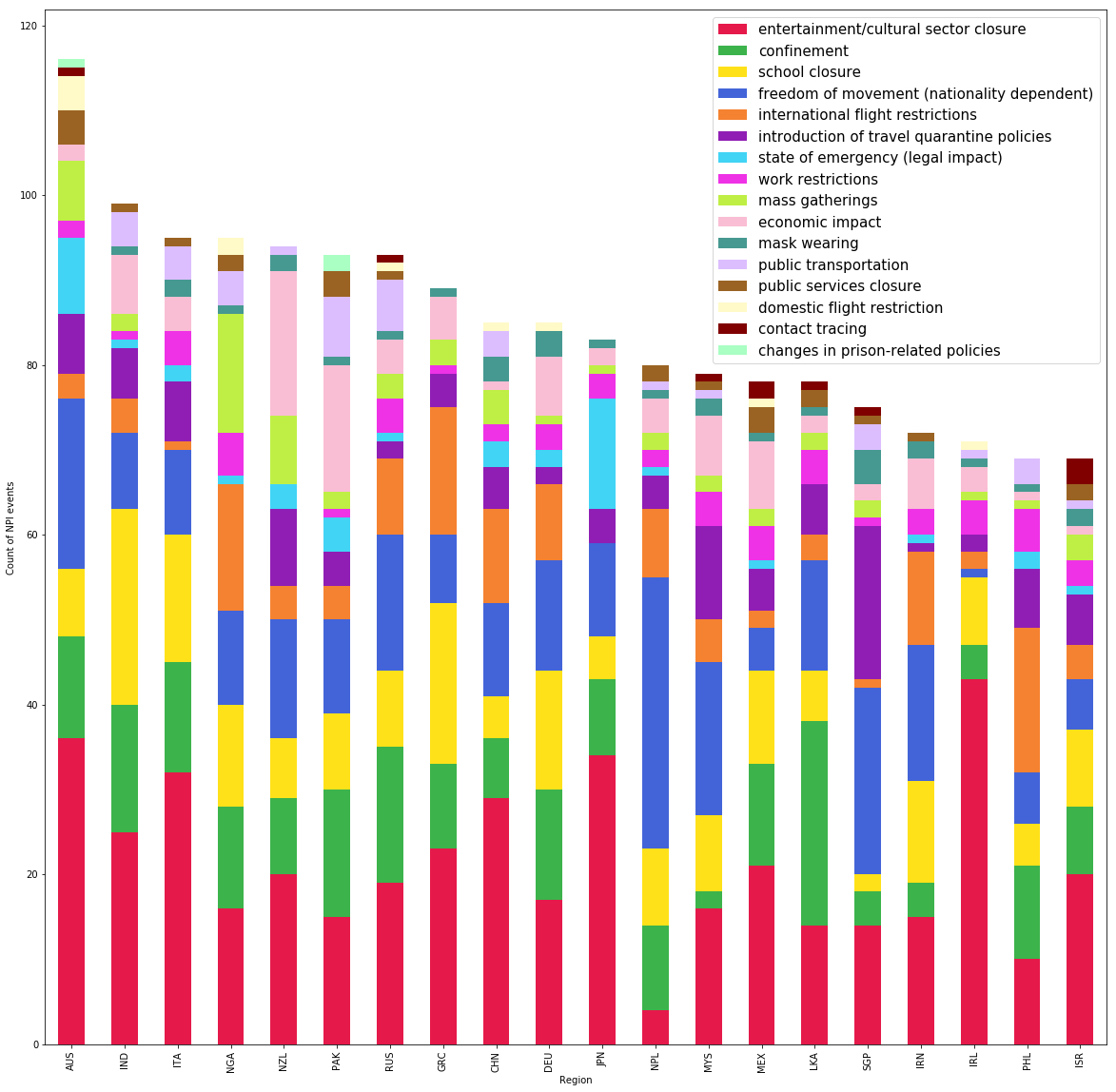}} \quad
    \subcaptionbox{}{\includegraphics[height=18\baselineskip]{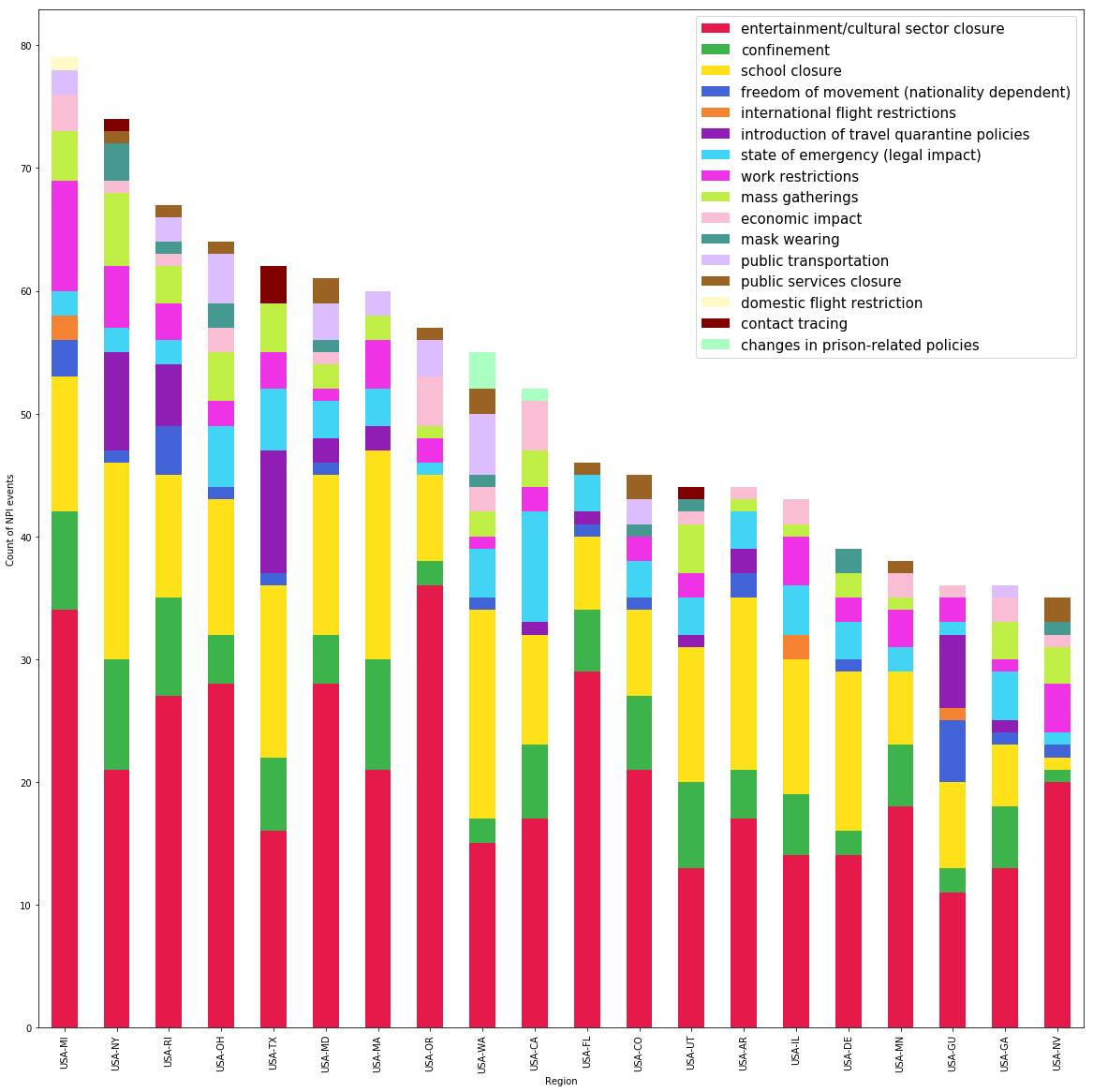}} \quad
    \caption{Distribution of \gls{npi} measures implemented in different geographies globally (left) and within US (right).} 
    \label{fig:stats-npi-count-by-region}
\end{figure}
 \begin{figure}[htp!]
    \centering
    \subcaptionbox{Travel-related \glspl{npi}\label{fig:travel-related}}{\includegraphics[height=12\baselineskip,width=.45\linewidth]{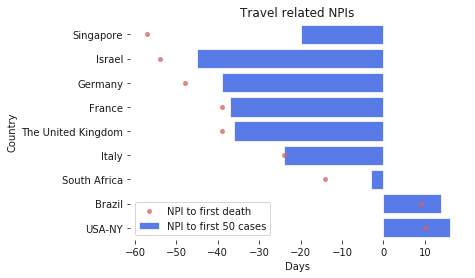}} \quad
    \subcaptionbox{Community-related \glspl{npi}\label{fig:community-related}}{\includegraphics[height=12\baselineskip,width=.45\linewidth]{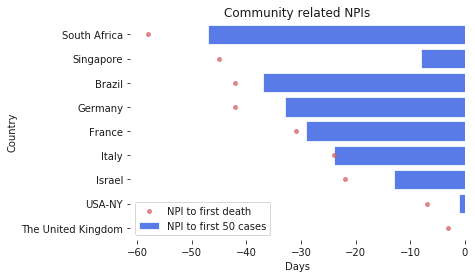}} \quad
    \caption{Elapsed time (in days) between the introduction of \glspl{npi} and recording of first death (\textcolor{red}{red}) or 50 cases (\textcolor{blue}{blue}) in countries that implemented travel-related vs community-related \glspl{npi} first.} 
    \label{fig:travel-community-related}
\end{figure}
 \begin{figure}[htp!]
    \centering
    \subcaptionbox{Imposed\label{fig:imposed_count_by_month}}{\includegraphics[height=12\baselineskip,width=.45\linewidth]{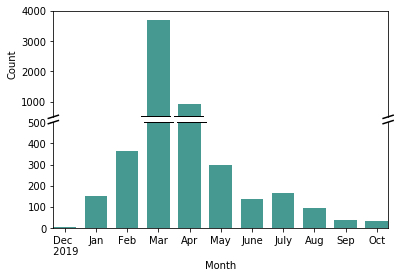}} \quad
    \subcaptionbox{Lifted\label{fig:lifted_count_by_month}}{\includegraphics[height=12\baselineskip,width=.45\linewidth]{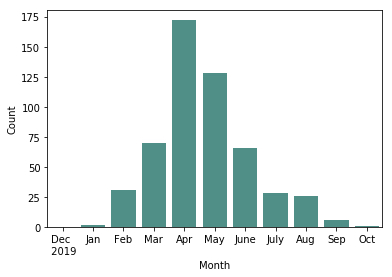}} \quad
    \caption{Number of imposed and lifted \glspl{npi} measures per month.} 
    \label{fig:number-imposed-lifted}
\end{figure}
 
\begin{figure}[htp!]
    \captionsetup[subfigure]{labelformat=empty}
    \centering
    \subcaptionbox{}{\includegraphics[height=10\baselineskip,width=.4\linewidth]{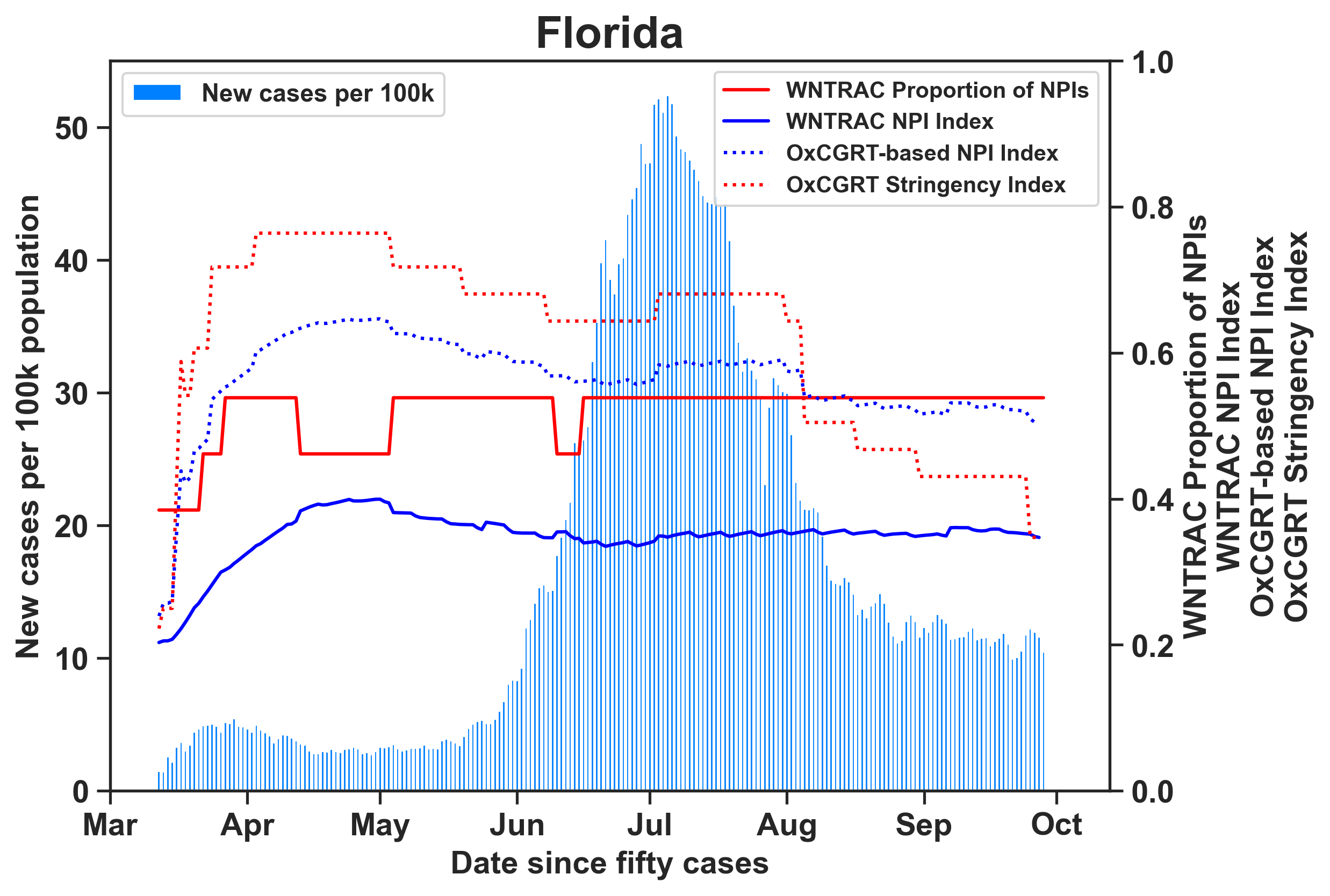}}\quad
    \subcaptionbox{}{\includegraphics[height=10\baselineskip,width=.4\linewidth]{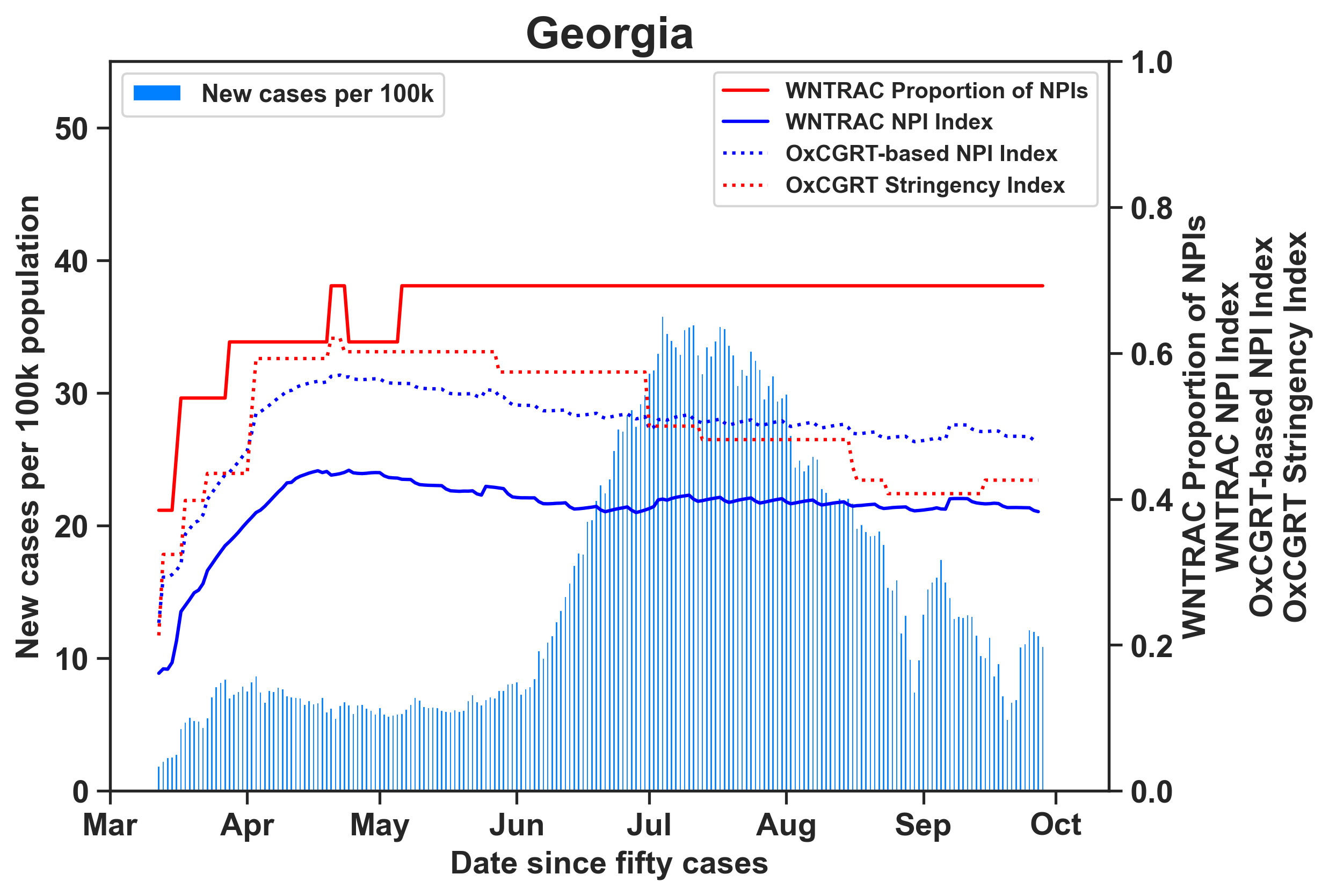}} \quad
    \subcaptionbox{}{\includegraphics[height=10\baselineskip,width=.4\linewidth]{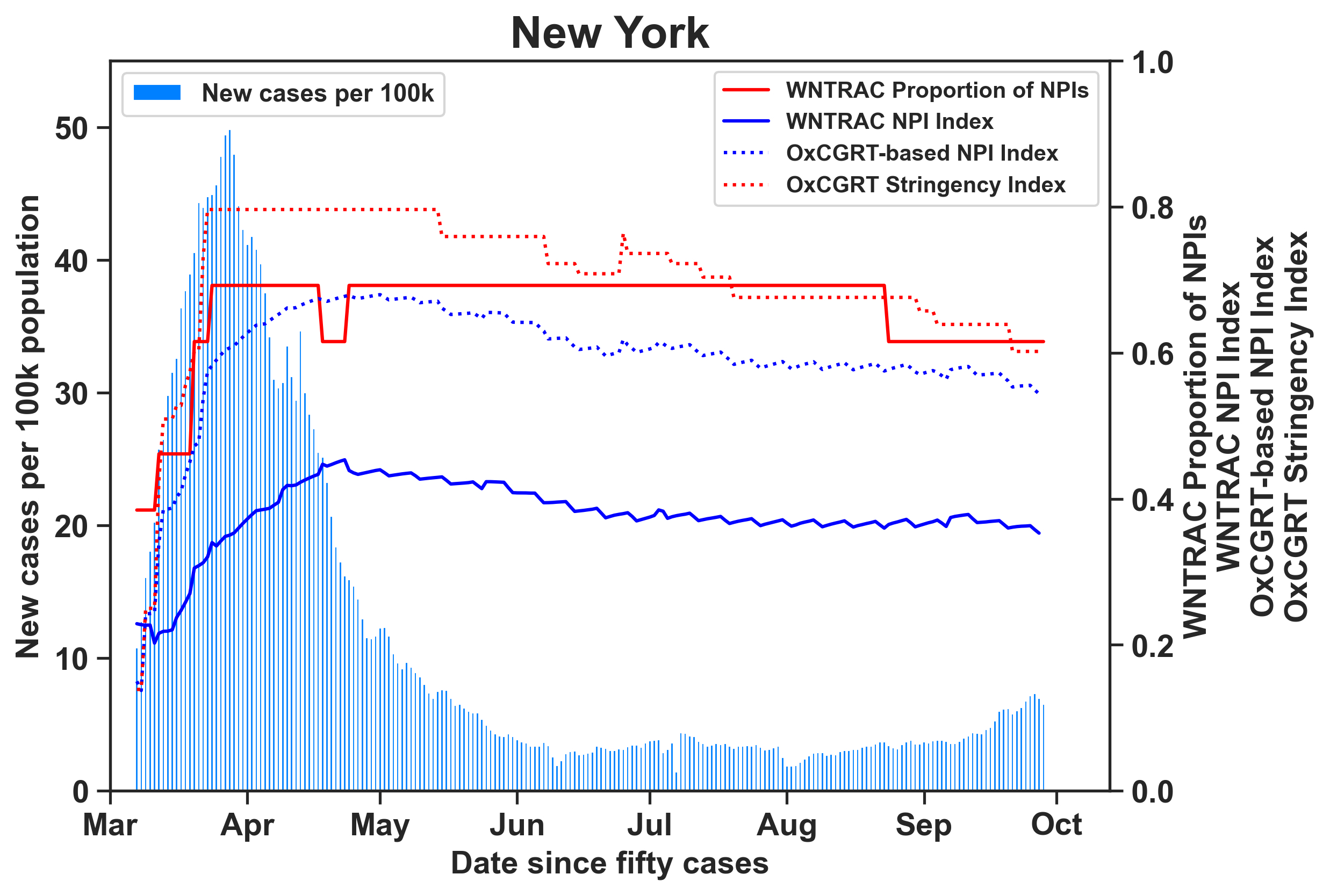}} \quad
    \subcaptionbox{}{\includegraphics[height=10\baselineskip,width=.4\linewidth]{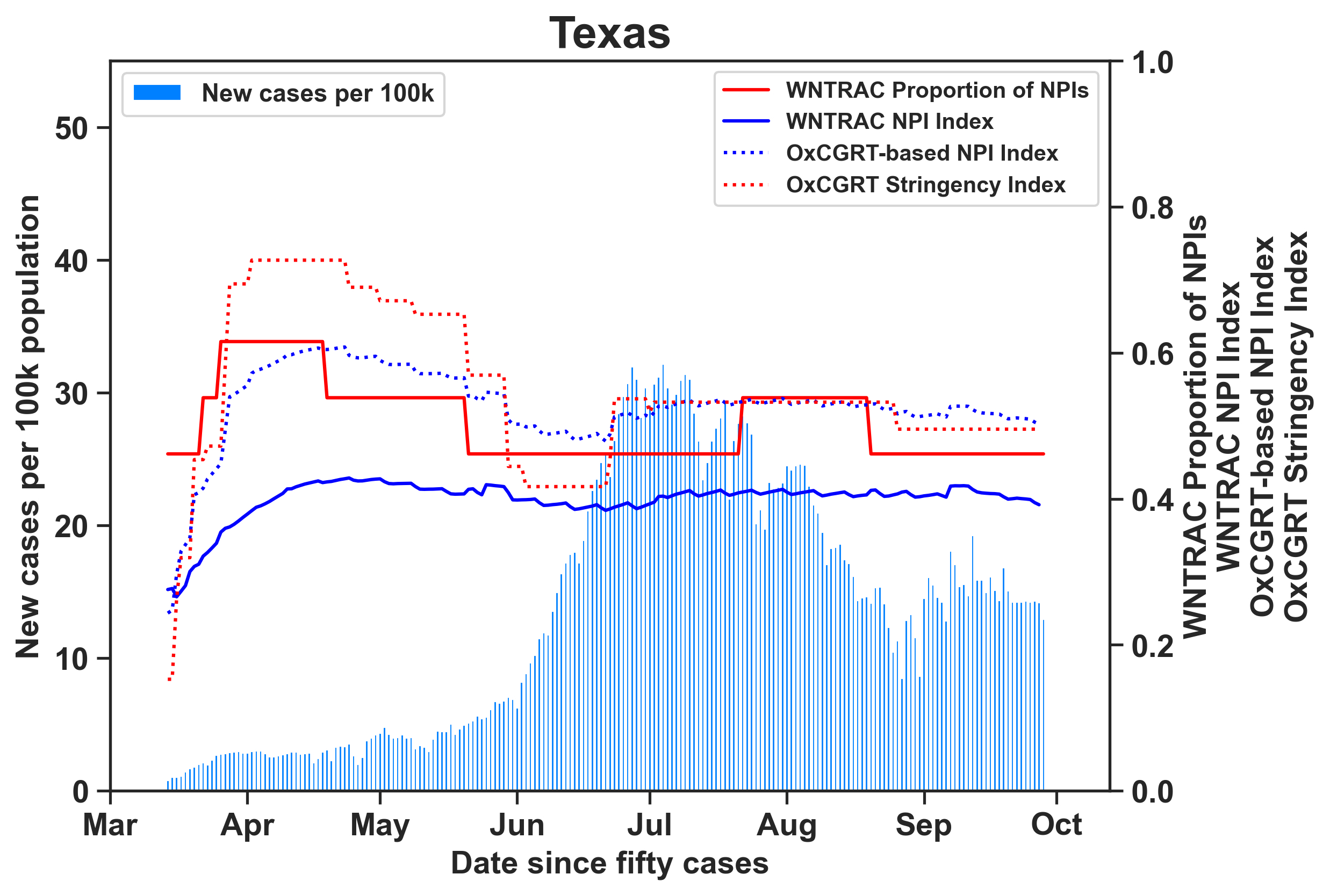}} \quad
    \caption{Trends in \gls{covid} cases per 100,1000 population and the \gls{npi}-based indices in representative US states.} 
    \label{fig:npi-trends}
\end{figure}

\clearpage
\newcommand{\pf}[1]{\parbox{6cm}{#1}}
\newcommand{\midsepdefault}{\aboverulesep = 0.605mm \belowrulesep = 0.984mm}
\newcommand{\midsepremove}{\aboverulesep = 0mm \belowrulesep = 0mm}

\begin{scriptsize}\centering
\begin{longtable}[t]{@{}p{0.15\textwidth}@{}p{0.35\textwidth}p{0.06\textwidth}@{}m{0.30\textwidth}}
        \toprule
        \textbf{\gls{npi}}
        & \textbf{Example} 
        & \textbf{Value} 
        & \textbf{Value description}\\
        \midrule
 
        {\parbox{2.5cm}{changes in \newline prison-related policies}} 
        & \pf{On March 30, the GNA announced the release of 466 detainees in Tripoli, as part of an effort to stop the spread of the virus in prisons.}
        & Integer
        & Number of prisoners that were released \\
        \midrule
        
        {confinement}
        &\pf{On 19 March, President Alberto Fernández announced a mandatory lockdown to curb the spread of coronavirus.}
        & Category
        & \begin{enumerate}[nosep, noitemsep, leftmargin=*]
            \item  Mandatory/advised for all the population
            \item  Mandatory/advised for people at risk
\end{enumerate} \\
        \midrule
        
        contact tracing
        & \pf{On 2 March, a case in Nimes was traced to the mid-February Mulhouse Megachurch event.}
        & Category 
        & \begin{enumerate}[nosep, noitemsep, leftmargin=*]
            \item Tracing back 14 days of contacts of a confirmed patient through electronic information
            \item Tracing contacts of a person who needs to be isolated as was in contact with a confirmed patient through electronic information
        \end{enumerate} \\
        \midrule
        
        {\parbox{2.5cm} {domestic flight restriction}}	
        & \pf{On 1 April, the Government of Afghanistan suspended flights between Kabul and Herat.}
        & String 
        & Name of the state where the passenger is arriving from \\
        \midrule
        
        economic impact 
        & \pf{Up until 14 March, the Afghan government had spent \$25 million to tackle the outbreak, which included \$7 million of aid packages.}
        & Category  
        & \begin{enumerate}[nosep, noitemsep,  leftmargin=*]
            \item  Stock market
            \item  Unemployment rate
            \item  Industrial production
\end{enumerate} \\
        \midrule
        
        {\parbox{2.5cm} {entertainment / \newline cultural sector closure}}
        & \pf{On April 7, Rockland and Sullivan counties closed their parks.}
        & Category  
        & \begin{enumerate}[nosep, noitemsep,  leftmargin=*]
              \item  Bars, restaurants, night clubs
              \item  Museums, theaters, cinema, libraries, festivities 
              \item  Parks and public gardens
              \item  Gyms and pools 
              \item  Churches
\end{enumerate} \\
        \midrule
        
        {\parbox{2.5cm} {freedom of movement \newline(nationality dependent)}}
        & \pf{Iran was added to the list of countries whose nationals were suspended entry to Cambodia, making a total of six.} 
        & String 
        & Name of the country the citizen is from\\
        \midrule

        {\parbox{2.5cm} {international \newline flight restrictions}}	
        & \pf{With effect from midnight on 1 April, Cuba suspended the arrival of all international flights.}
        & String 
        & Name of the country or state where the passenger is arriving from \\
        \midrule
        
        {\parbox{2.5cm} {introduction of \newline travel quarantine policies}}	
        & \pf{Israeli nationals returning from Egypt were required to enter an immediate 14-day quarantine.}
        & String 
        & Name of the country or state where the passenger travelled from\\
        \midrule
        
        mask wearing 
        & \pf{On April 15, Cuomo signed an executive order requiring all New York State residents to wear face masks or coverings in public places.}
        & Category 
        & \begin{enumerate}[nosep, noitemsep,  leftmargin=*, topsep=0pt] \item Mandatory
            \item Mandatory in some public spaces
            \item Recommended
\end{enumerate} \\
        \midrule
        
        mass gatherings	
        & \pf{On 13 March, it was announced at an official press conference that a four-week ban on public gatherings of more than 100 persons would be put into effect as of Monday 16 March.}
        & Integer 
        & Maximum number of people in social gatherings allowed by the government\\
        \midrule
        
        public services closure
        & \pf{On 19 March, Election Commissioner Mahinda Deshapriya revealed that the 2020 Sri Lankan parliamentary election will be postponed indefinitely until further notice due to the coronavirus pandemic.} 
        & Category 
        & \begin{enumerate}[nosep, noitemsep,  leftmargin=*]
            \item Government/parliament system closed
            \item Legal system closed
\end{enumerate} \\  
        \midrule
        
        public transportation
        & \pf{On March 20, Regina Transit and Saskatoon Transit suspended fares for all bus service, but with reduced service.} 
        & Category 
        &  \begin{enumerate}[nosep, noitemsep,  leftmargin=*]
            \item Partial cancellation of routes/stops during the week/weekend
            \item Total cancellation of transport (special case for some states in China)
\end{enumerate} \\   
        \midrule
        
         school closure
         & \pf{On 13 March, the Punjab and Chhattisgarh governments declared holidays in all schools and colleges till 31 March.}
         & Category 
         & \begin{enumerate}[nosep, noitemsep,  leftmargin=*]
             \item All schools (general) closed
             \item Only kindergartens/daycare closed
             \item Only schools (primary/secondary) closed
             \item Universities closed
\end{enumerate} \\
        \midrule
        
        state of emergency \newline (legal impact)
        & {\pf{Governor Charlie Baker declared a state of emergency for the state of Massachusetts on March 10.}}
        & Category
        & \begin{enumerate}[nosep, noitemsep,  leftmargin=*]
            \item National guard joins the law enforcement
            \item Army joins the law enforcement
\end{enumerate} \\
        \midrule
        
        \renewcommand{\arraystretch}{1.5}work restrictions
        & \pf{On 10 April, Koike announced closure requests for six categories of businesses in Tokyo.}
        & Category 
        & \begin{enumerate}[nosep, noitemsep,  leftmargin=*]
            \item Suggestion to work from home for non-essential workers
            \item Mandatory work from home enforcement for non-essential workers 
\end{enumerate} \\
        \bottomrule
  \caption{Taxonomy of the \acrlong{wntrac} dataset.}\label{tab:taxonomy} 
  \end{longtable}
\end{scriptsize}

\begin{table*}[htp!]
    \scriptsize
\centering
    \begin{tabular}[t]{p{0.10\textwidth}p{0.50\textwidth}p{0.30\textwidth}}\toprule
            \textbf{Field name}
            & \textbf{Description} 
            & \textbf{Example}
            \\ \midrule
            even\_id & Globally unique identifier~\cite{wiki:uuid} for the particular \gls{npi} & 7db34fd1-d121-479f-9713-af7596a45aa1 \\
            type    &  Type of the \gls{npi} & School closure \\ 
	        country &  Country where the \gls{npi} was implemented. Name in ISO 3166-1 coding~\cite{wiki:iso-3166-1} & USA \\
            state/province &  State or province where the \gls{npi} was implemented. Name in ISO 3166-2 coding~\cite{wiki:iso-3166-2} & Vermont \\
            date & Date when the \gls{npi} comes to effect. It is not the date of announcement & 2020-03-26 \\
            epoch & Unix epoch time~\cite{wiki:epoch} corresponding to the date & 1589749200000.0 \\ 
            value & Value associated with the \gls{npi}. & Refer to Table for details \\
            restriction & Ordinal values representing imposition ($1$) or lifting ($0$) of an \gls{npi} & 0 \\
            sent\_id & Globally unique identifier~\cite{wiki:uuid} for the evidence sentence & d68ea644-24d5-4abf-93b0-dabc1cd3c2eb \\
            doc\_url & Document URL & \url{https://en.wikipedia.org/wiki/COVID-19_pandemic_in_Vermont} \\
            crawl\_id & Globally unique identifier~\cite{wiki:uuid} for the particular crawl in which this evidence sentence was fetched & 2020-05-06\_d0cba9ae-8fda-11ea-b351-069b8ffc8dc8 \\
            crawl\_date & Date of the crawl that fetched this evidence sentence &  2020\-05\-06 \\
            text & Evidence sentence in the document where the \gls{npi} is discussed & On March 26, Governor Scott ordered all schools in Vermont to remain closed for in-person classes for the rest of the academic year \\
            citation\_url & URL cited for the evidence sentence in the source document & \iffalse \url{https://governor.vermont.gov/content/directive-5-continuity-learning-planning-pursuant-eo-01-20} \fi \\
            anno\_provided\_url & Additional citation URL provided by the human volunteer who performed the validation. & \iffalse \url{https://www.vpr.org/post/gov-closes-vermont-schools-rest-academic-year} \fi \\
            fine\_grained\_location & Geographic locations mentioned in the evidence sentence separated by pipeline. & Vermont \\
            source\_type & Wikipedia citation source type indicating government ($G$) or other sources ($O$) & G \\
             \bottomrule
    \end{tabular}
    \caption{Data record for the \acrlong{wntrac} dataset.} \label{tab:data-record}
\end{table*} \begin{table*}[!htb]
    \centering
    \begin{tabular}{lcccccc}
        \toprule
        \multirow{2}{*}{} & \multicolumn{3}{c}{\textbf{All \gls{npi} event types}} & \multicolumn{3}{c}{\textbf{Top 5 \gls{npi} event types}} \\ \cmidrule{2-7} 
         & \textbf{A vs E\textsubscript{1}} & \textbf{A vs E\textsubscript{2}} & \textbf{E\textsubscript{1} vs E\textsubscript{2}} & \textbf{A vs E\textsubscript{1}} & \textbf{A vs E\textsubscript{2}} & \textbf{E\textsubscript{1} vs E\textsubscript{2}} \\
        \midrule
        \textbf{Type} & 0.63 & 0.69 & 0.80 & 0.81 & 0.77 & 0.85 \\ 
        \textbf{Type + Value} & 0.41 & 0.42 & 0.69 & 0.51 & 0.47 & 0.70 \\ 
        \textbf{Date} & 0.50 & 0.61 & 0.73 & 0.60 & 0.69 & 0.76 \\ 
        \textbf{Region} & 0.99 & 1.00 & 0.99 & 0.98 & 1.00 & 0.98 \\ 
        \textbf{Restriction} & 0.36 & 0.43 & 0.74 & 0.74 & 0.58 & 0.69 \\
        \textbf{Type + Date} & 0.44 & 0.53 & 0.70 & 0.51 & 0.59 & 0.72 \\ 
        \textbf{Type + Value + Date} & 0.31 & 0.34 & 0.62 & 0.35 & 0.36 & 0.59 \\ 
        \textbf{Type + Value + Date + Region} & 0.30 & 0.33 & 0.62 & 0.35 & 0.36 & 0.59 \\ 
        \textbf{Type + Value + Date + Region + Restriction} & 0.26 & 0.29 & 0.61 & 0.34 & 0.35 & 0.59 \\ \bottomrule
    \end{tabular}
    \caption{Inter-annotator agreement between average volunteers (A) and two groups of experienced volunteers (E\textsubscript{1} and E\textsubscript{2}). Region includes both country and state/territories as applicable.}
    \label{tab:iaa}
\end{table*} \clearpage
\begin{table}[!htb]
    \begin{minipage}{.33\linewidth}
      \centering
      \scriptsize
        \begin{tabular}{ll}
        \textbf{Region} & \textbf{Code} \\ \midrule
        Abkhazia & GEO-AB \\
        Afghanistan & AFG \\
        Albania & ALB \\
        Algeria & DZA \\
        Andorra & AND \\
        Angola & AGO \\
        Anguilla & AIA \\
        Antarctica & ATA \\
        Antigua and Barbuda & ATG \\
        Argentina & ARG \\
        Armenia & ARM \\
        Australia & AUS \\
        Austria & AUT \\
        Azerbaijan & AZE \\
        Bahamas & BHS \\
        Bahrain & BHR \\
        Bangladesh & BGD \\
        Barbados & BRB \\
        Belarus & BLR \\
        Belgium & BEL \\
        Belize & BLZ \\
        Benin & BEN \\
        Bhutan & BTN \\
        Bolivia & BOL \\
        Bosnia and Herzegovina & BIH \\
        Botswana & BWA \\
        Brazil & BRA \\
        Brunei & BRN \\
        Bulgaria & BGR \\
        Burkina Faso & BFA \\
        Burundi & BDI \\
        Cambodia & KHM \\
        Cameroon & CMR \\
        Canada & CAN \\
        Cape Verde & CPV \\
        Central African Republic & CAF \\
        Chad & TCD \\
        Chile & CHL \\
        China & CHN-TB \\
        Colombia & COL \\
        Comoros & COM \\
        Costa Rica & CRI \\
        Croatia & HRV \\
        Cuba & CUB \\
        Cyprus & CYP \\
        Czech Republic & CZE \\
        Democratic Republic of Congo & COD \\
        Denmark & DNK \\
        Djibouti & DJI \\
        Dominica & DMA \\
        Dominican Republic & DOM \\
        Ecuador & ECU \\
        Egypt & EGY \\
        El Salvador & SLV \\
        Equatorial Guinea & GNQ \\
        Eritrea & ERI \\
        Estonia & EST \\
        Eswatini & SWZ \\
        Ethiopia & ETH \\
        Faroe Islands & DEN-FI \\
        Fiji & FJI \\
        Finland & FIN \\
        France & FRA \\
        French Guiana & GUF \\
        French Polynesia & PYF \\
        French Saint Martin & MAF \\
        Gabon & GAB \\
        Gambia & GMB \\
        Georgia (country) & GEO \\
        \end{tabular}
    \end{minipage}\begin{minipage}{.33\linewidth}
      \centering
      \scriptsize
        \begin{tabular}{ll}
        Germany & DEU \\
        Ghana & GHA \\
        Greece & GRC \\
        Greenland & DEN-GR \\
        Grenada & GRD \\
        Guadeloupe & FRA-GU \\
        Guatemala & GTM \\
        Guinea & GIN \\
        Guinea-Bissau & GNB \\
        Guyana & GUY \\
        Haiti & HTI \\
        Honduras & HND \\
        Hong Kong & CHN-HK \\
        Hungary & HUN \\
        Iceland & ISL \\
        India & IND \\
        Indonesia & IDN \\
        Iran & IRN \\
        Iraq & IRQ \\
        Israel & ISR \\
        Italy & ITA \\
        Ivory Coast & CIV \\
        Jamaica & JAM \\
        Japan & JPN \\
        Jordan & JOR \\
        Kazakhstan & KAZ \\
        Kenya & KEN \\
        Kosovo & KOS \\
        Kuwait & KWT \\
        Kyrgyzstan & KGZ \\
        Laos & LAO \\
        Latvia & LVA \\
        Lebanon & LBN \\
        Lesotho & LSO \\
        Liberia & LBR \\
        Libya & LBY \\
        Liechtenstein & LIE \\
        Lithuania & LTU \\
        Luhansk People's Republic & UKR-09 \\
        Luxembourg & LUX \\
        Madagascar & MDG \\
        Mainland China & CHN \\
        Malawi & MWI \\
        Malaysia & MYS \\
        Maldives & MDV \\
        Mali & MLI \\
        Malta & MLT \\
        Martinique & FRA-MA \\
        Mauritania & MRT \\
        Mauritius & MUS \\
        Mayotte & FRA-MA \\
        Mexico & MEX \\
        Moldova & MDA \\
        Monaco & MCO \\
        Mongolia & MNG \\
        Montenegro & MNE \\
        Morocco & MAR \\
        Mozambique & MOZ \\
        Myanmar & MMR \\
        Namibia & NAM \\
        Nepal & NPL \\
        Netherlands & NLD \\
        New Caledonia & FRA-NC \\
        New Zealand & NZL \\
        Nicaragua & NIC \\
        Niger & NER \\
        Nigeria & NGA \\
        North Korea & PRK \\
        North Macedonia & MKD \\
        Northern Cyprus & CYP \\
        \end{tabular}
    \end{minipage} 
    \begin{minipage}{.33\linewidth}
      \scriptsize
        \begin{tabular}{ll}
        Norway & NOR \\
        Oman & OMN \\
        Pakistan & PAK \\
        Panama & PAN \\
        Papua New Guinea & PNG \\
        Paraguay & PRY \\
        Peru & PER \\
        Philippines & PHL \\
        Poland & POL \\
        Portugal & PRT \\
        Qatar & QAT \\
        Republic of Artsakh & ARM \\
        Republic of Congo & COG \\
        Republic of Ireland & IRL \\
        Romania & ROU \\
        Russia & RUS \\
        Rwanda & RWA \\
        Saint Helena & SHN \\
        Saint Kitts and Nevis & KNA \\
        Saint Lucia & LCA \\
        Saint Pierre and Miquelon & SPM \\
        Saint Vincent and Grenadines & VCT \\
        San Marino & SMR \\
        Saudi Arabia & SAU \\
        Senegal & SEN \\
        Serbia & SRB \\
        Seychelles & SYC \\
        Sierra Leone & SLE \\
        Singapore & SGP \\
        Slovakia & SVK \\
        Slovenia & SVN \\
        Somalia & SOM \\
        Somalia & SOM \\
        South Africa & ZAF \\
        South Korea & KOR \\
        South Sudan & SSD \\
        Spain & ESP \\
        Sri Lanka & LKA \\
        State of Palestine & PSE \\
        Sudan & SDN \\
        Suriname & SUR \\
        Sweden & SWE \\
        Switzerland & CHE \\
        Syria & SYR \\
        Taiwan & TWN \\
        Tajikistan & TJK \\
        Tanzania & TZA \\
        Thailand & THA \\
        Timor-Leste & TLS \\
        Togo & TGO \\
        Trinidad and Tobago & TTO \\
        Tunisia & TUN \\
        Turkey & TUR \\
        Turkmenistan & TKM \\
        Uganda & UGA \\
        United Arab Emirates & ARE \\
        United Kingdom & GBR \\
        United States & USA \\
        Uruguay & URY \\
        Uzbekistan & UZB \\
        Vatican City & VAT \\
        Venezuela & VEN \\
        Vietnam & VNM \\
        Western Sahara & ESH \\
        Yemen & YEM \\
        Zambia & ZMB \\
        Zimbabwe & ZWE \\
                 & \\
                 & \\
                 & \\
        \end{tabular}
    \end{minipage} 
    \caption{List of regions currently supported by the \gls{wntrac} dataset.}
    \label{tab:regions-global}
\end{table}

\clearpage

\begin{table}[!htb]
    \begin{minipage}{.5\linewidth}
      \centering
      \scriptsize
        \begin{tabular}{ll}
        \textbf{Region} & \textbf{Code} \\ \midrule
        Alabama & USA-AL \\
        Alaska & USA-AK \\
        Arizona & USA-AZ \\
        Arkansas & USA-AR \\
        California & USA-CA \\
        Colorado & USA-CO \\
        Connecticut & USA-CT \\
        Delaware & USA-DE \\
        Florida & USA-FL \\
        Georgia (U.S. state) & USA-GA \\
        Guam & USA-GU \\
        Guantanamo Bay Naval Base & USA-Guantanamo\_Bay\_Naval\_Base \\
        Hawaii & USA-HI \\
        Idaho & USA-ID \\
        Illinois & USA-IL \\
        Indiana & USA-IN \\
        Iowa & USA-IA \\
        Kansas & USA-KS \\
        Kentucky & USA-KY \\
        Louisiana & USA-LA \\
        Maine & USA-ME \\
        Maryland & USA-MD \\
        Massachusetts & USA-MA \\
        Michigan & USA-MI \\
        Minnesota & USA-MN \\
        Mississippi & USA-MS \\
        Missouri & USA-MO \\
        Montana & USA-MT \\
        \end{tabular}
    \end{minipage}\begin{minipage}{.5\linewidth}
      \centering
      \scriptsize
        \begin{tabular}{ll}
        Nevada & USA-NV \\
        New Hampshire & USA-NH \\
        New Jersey & USA-NJ \\
        New Mexico & USA-NM \\
        New York (state) & USA-NY \\
        North Carolina & USA-NC \\
        North Dakota & USA-ND \\
        Northern Mariana Islands & USA-Northern\_Mariana\_Islands \\
        Ohio & USA-OH \\
        Oklahoma & USA-OK \\
        Oregon & USA-OR \\
        Pennsylvania & USA-PA \\
        Puerto Rico & USA-PR \\
        Rhode Island & USA-RI \\
        South Carolina & USA-SC \\
        South Dakota & USA-SD \\
        Tennessee & USA-TN \\
        Texas & USA-TX \\
        United States Virgin Islands & USA-Virgin\_Islands \\
        Utah & USA-UT \\
        Vermont & USA-VT \\
        Virginia & USA-VA \\
        Washington (state) & USA-WA \\
        Washington D.C. & USA-DC \\
        West Virginia & USA-WV \\
        Wisconsin & USA-WI \\
        Wyoming & USA-WY \\
        \end{tabular}
    \end{minipage} 
    \caption{List of US states and territories currently supported by the \gls{wntrac} dataset.}
    \label{tab:regions-us}
\end{table}  }{}
\end{document}